\documentclass[a4paper,fleqn,usenatbib]{mnras}

\usepackage{makecell}
\usepackage[T1]{fontenc}
\usepackage{ae,aecompl}

\usepackage{graphicx}	% Including figure files
\usepackage{amsmath}	% Advanced maths commands
\usepackage{amssymb}	% Extra maths symbols

\usepackage{abbrevs}
\usepackage{xspace}
\usepackage[dvipsnames]{xcolor}

\begingroup
\catcode`\_=\active
\gdef_#1{\ensuremath{\sb{\mathrm{#1}}}}
\endgroup
\mathcode`\_=\string"8000
\catcode`\_=12

\newcommand{\rev}{\textcolor{Black}}

\newcommand{\Msolar}{M$_{\odot}$\xspace}
\newcommand{\Msolarpc}{M$_{\odot}$ / pc$^{2}$\xspace}

\newcommand{\atcc}{cm$^{-3}$\xspace}

\newcommand{\linexy}[2]{\ifmmode \mathrm{#1{\sc #2}} \else #1{\sc #2}\xspace \fi}
\def\hii{\linexy{H}{ii}}

\newabbrev\ISM{interstellar medium (ISM)}[ISM]
\newabbrev\CSM{circumstellar medium (CSM)}[CSM]
\newabbrev\WNM{Warm Neutral Medium (WNM)}[WNM]
\newabbrev\WIM{Warm Ionised Medium (WIM)}[WIM]
\newabbrev\CNM{Cold Neutral Medium (CNM)}[CNM]
\newabbrev\IMF{Initial Mass Function (IMF)}[IMF]
\newabbrev\CMF{Core Mass Function (IMF)}[IMF]
\newabbrev\AMR{Adaptive Mesh Refinement (AMR)}[AMR]
\newabbrev\HGB{Horizontal Giant Branch (HGB)}[HGB]
\newabbrev\SFE{Star Formation Efficiency (SFE)}[SFE]
\newabbrev\TSFE{Total Star Formation Efficiency (TSFE)}[TSFE]
\newabbrev\OSFE{Observed Star Formation Efficiency (OSFE)}[OSFE]
\newabbrev\SFR{Star Formation Rate (SFR)}[SFR]
\newabbrev\YSOs{Young Stellar Objects (YSOs)}[YSOs]
\newabbrev\YSO{Young Stellar Object (YSO)}[YSO]
\newabbrev\PDF{Probability Distribution Function (PDF)}[PDF]
\newabbrev\PSF{Point Spread Function (PSF)}[PSF]
\newabbrev\LMC{Large Magellanic Cloud (LMC)}[LMC]
\newabbrev\EUV{extreme ultraviolet radiation (EUV)}[EUV]
\newabbrev\SIS{singlular isothermal sphere (SIS)}[SIS]

\makeatletter
\newcommand*\bigcdot{\mathpalette\bigcdot@{.5}}
\newcommand*\bigcdot@[2]{\mathbin{\vcenter{\hbox{\scalebox{#2}{$\m@th#1\bullet$}}}}}
\makeatother

\makeatletter
\renewcommand\maybe@space@{%
	\maybe@ictrue % <= this is new
	\expandafter   \@tfor
	\expandafter \reserved@a
	\expandafter :%
	\expandafter =%
	\nospacelist
	\do \t@st@ic
	\ifmaybe@ic % <= this is new
	\space
	\fi
}
\makeatother
\title[When \hii Regions are Complicated]{When \hii Regions are Complicated: Considering Perturbations from Winds, Radiation Pressure, and Other Effects}

\author[Geen et al]{
Sam Geen,$^{1,2}$\thanks{E-mail: sam.geen@uni-heidelberg.de}
Eric Pellegrini,$^{1}$
Rebekka Bieri,$^{3}$
Ralf Klessen$^{1,4}$
\\
{$^{1}$ Universit\"at Heidelberg, Zentrum f\"ur Astronomie, Institut f\"ur Theoretische Astrophysik, Albert-Ueberle-Str. 2, 69120 Heidelberg, Germany}\\
{$^{2}$ Anton Pannekoek Institute for Astronomy, Universiteit van Amsterdam, Science Park 904, 1098 XH Amsterdam, The Netherlands}\\
{$^{3}$ Max-Planck-Institute for Astrophysics, Karl-Schwartzschild-Strasse 1, Garching, Germany}\\
{$^{4}$ Universit\"at Heidelberg, Interdisziplin\"ares Zentrum fur Wissenschaftliches Rechnen, INF 205, 69120, Heidelberg, Germany}\\
}

\date{\today}

\begin{document}
\label{firstpage}
\pagerange{\pageref{firstpage}--\pageref{lastpage}}
\maketitle

\begin{abstract}
\rev{We explore to what extent simple algebraic models can be used to describe \hii regions when winds, radiation pressure, gravity and photon breakout are included. We a) develop algebraic models to describe the expansion of photoionised \hii regions under the influence of gravity and accretion in power-law density fields with $\rho \propto r^{-w}$, b) determine when terms describing winds, radiation pressure, gravity and photon breakout become significant enough to affect the dynamics of the \hii region where $w=2$, and c) solve these expressions for a set of physically-motivated conditions. We find that photoionisation feedback from massive stars is the principal mode of feedback on molecular cloud scales, driving accelerating outflows from molecular clouds in cases where the peaked density structure around young massive stars is considered at radii between $\sim$0.1 and 10-100 pc. Under a large range of conditions the effect of winds and radiation on the dynamics of \hii regions is around 10\% of the contribution from photoionisation. The effect of winds and radiation pressure are most important at high densities, either close to the star or in very dense clouds such as those in the Central Molecular Zone of the Milky Way. Out to $\sim$0.1 pc they are the principal drivers of the \hii region. Lower metallicities make the relative effect of photoionisation even stronger as the ionised gas temperature is higher.} 
\end{abstract}


%%%%%%%%%%%%%%%%%%%%%%%%%%%%%%%%%%%%%%%%%%%%%%%%%%

%%%%%%%%%%%%%%%%% BODY OF PAPER %%%%%%%%%%%%%%%%%%

\section{Introduction}
\label{introduction}

Massive stars produce large quantities of energy as radiation and kinetic outflows that shape astrophysical phenomena on a wide range of scales. For a holistic view of this process of stellar ``feedback'', we must understand the behaviour of such flows on a systemic level. In this paper, we produce new models to describe the evolution of feedback structures around massive stars in their natal environment, \hii regions. These are important in understanding how kinetic energy and radiation from massive stars is transmitted from the sub-parsec scale to the galactic scale.

\hii regions are volumes of ionised hydrogen around massive stars. Their evolution is a complex process, affected by stellar evolution, gas dynamics and the interaction of different feedback processes. The complexity of the problem creates a large parameter space to study.

Recently, numerical simulations have been used to model \hii regions with increasing physical fidelity, but they are expensive and do not offer descriptive explanations for why a certain set of model assumptions leads to a particular outcome. Algebraic expressions that describe \hii regions, while much simplified, allow faster parameter space exploration and quantitative predictions. A large body of analytic work exists to describe individual processes that shape \hii regions, and to some extent how they interact, though this work is often limited in its predictive power by (necessary) simplifications made in describing the modelled systems (see references in Section \ref{introduction:hiiregions} to \ref{introduction:interactingprocesses}).

This work brings three new insights to the problem. Firstly, we provide a new dynamical equation to describe the evolution of photoionised \hii regions from young massive stars sitting inside physically-motivated power law gas density profiles \rev{($\rho \propto r^{-w}$)} with gravity and accretion. This equation behaves differently to previous formulations of the expansion in uniform environments.

\rev{Secondly, we provide formulations for how winds, radiation pressure, photon breakout and self-gravity influence these \hii regions when $w=2$.}

Thirdly, we calculate surfaces in phase space where winds, radiation pressure, gravity and photon breakout significantly shape the evolution of \hii regions in physically-motivated conditions. With this, we seek to adress the question of when \hii regions are \textit{complicated}, i.e. where we must use more detailed numerical solutions \rev{than our simple algebraic models} to explain their evolution.

We begin this Section by summarising existing analytic work on the evolution of \hii regions. We describe theory for how the photoionisation of H to H$^{+}$ leads to an overpressurised, expanding structure. We then discuss theory of stellar winds and direct radiation pressure, and study how these processes interact with the photoionised \hii region. Finally, we lay out the structure of this paper and how we address this problem.

\subsection{Photoionised \hii Regions}
\label{introduction:hiiregions}

The evolution of \hii regions has been studied widely using observations, analytic theory, and more recently, hydrodynamical simulations. Theories governing their expansion have been proposed by several authors since the 1950s. Hydrogen is ionised by photons with energies above 13.6 eV, while dust and other species can be ionised by lower energy photons. \cite{KahnF.D.1954} derived plane-parallel equations to describe ionisation fronts. They defined a transition between R-type fronts, where the front is bounded by the ability for new ionising photons to reach neutral gas, and D-type fronts, which are in photoionisation equilibrium and expand because they are warmer and therefore overpressured compare to the surrounding medium (approx. $10^4$ K versus 10 to 100 K). In a uniform medium, the ionisation front typically begins in the R-type phase and over approximately one recombination time (the amount of time it takes for the ionised gas to recombine to its neutral state) it transitions to the D-type phase. This time is typically short in dense cloud environments \cite[see our previous analysis in][]{Geen2015a}.

Spherically-symmetric solutions were derived by authors such as \cite{SpitzerLyman1978}, \cite{Dyson1980} and \cite{Franco1990}, while \cite{Whitworth1979} and \cite{Franco1989} describe ``blister'' \hii regions located close to the cloud surface and able to escape preferentially from the cloud in one direction. More recently, \cite{Hosokawa2006} and \cite{Raga2012} modified these equations to consider the inertia of the shell and external pressure terms from the neutral gas. These results have been used to test simulation codes by \cite{Bisbas2015} and in various more physically complete models. 

On small scales, \cite{Keto2002} and \cite{Keto2007} model Ultracompact \hii regions whose expansion is resisted by the gravity of, and accretion onto, a massive protostar. \cite{Shapiro2006} describe relativistic ionisation fronts in cosmological conditions, while \cite{Alvarez2006} derive a condition for \hii regions to overflow the D-type front based on the solutions of \cite{Shu2002}, which has been explored in nearby galaxies by \cite{Seon2009} and \cite{Pellegrini2012}. The solutions given in these papers were further extended and applied by \cite{Tremblin2014a}, \cite{Didelon2015} and \cite{Geen2015b} to describe the evolution of \hii regions in observed and simulated molecular clouds, assuming different external density profiles. In Section \ref{photoionisation_only}, we use these model assumptions to derive a dynamical equation for how gravity and accretion shape \hii regions in power-law density fields found around young massive stars in clumps and molecular clouds. 

See the review by \cite{Dale2015a} or the introduction of \cite{Geen2018} for a more detailed discussion on 3D numerical simulations of photoionised \hii regions, which are beyond the scope of this paper.

\subsection{Stellar Winds}
\label{introduction:winds}

Stellar winds ejected from the surface of the star directly deposit kinetic energy into the \CSM. Early work by \cite{Avedisova1972}, \cite{Castor1975} and \cite{Weaver1977} demonstrated that in the absence of radiative cooling, the cumulative energy emitted by a star over its lifetime can drive the expansion of hot bubbles around the host star(s). This is backed up by more recent work such as \cite{Fierlinger2016}. Observational evidence for wind-driven bubbles is given by hot X-ray emission \citep[e.g][]{Dunne2003}, which requires hot gas above $10^6$ K shocked by highly energetic kinetic flows. However, if most of this energy is lost to radiative cooling \citep[see][]{MacLow1988}, the influence of winds is strongly diminished. 

The transition from purely adiabatic to efficiently cooling winds is discussed in the context of \hii regions by \cite{Rahner2017}, \cite{Silich2018} and \cite{Rahner2019}, where the cooling rate of winds greatly affects their ability to drive an expanding shell. \cite{Silich2017} additionally argue that if the wind bubbles around individual stars do not merge, the combined effect of winds is further reduced. \cite{Harper-Clark2009} and \cite{Krumholz2009} also argue that winds can easily escape a porous shell owing to their high velocities. This causes a further reduction in the ability for winds to drive a spherical shell outwards.

\subsection{Radiation Pressure}
\label{introduction:radiationpressure}

Radiation pressure from photons of various energies has been proposed as a mechanism for driving flows in astrophysical environments. \cite{Mathews1967}, \cite{Krumholz2009}, \cite{Murray2010}, \cite{Draine2011} and \cite{Kim2016} have highlighted the role of radiation pressure in driving the expansion of \hii regions. The structure of \hii regions with strong sources of radiation pressure is complex. \cite{Pellegrini2007} and \cite{Pellegrini2009} argue that radiation pressure sets up a density gradient in \hii regions due to the differential flux of ionising photons throughout the \hii region.

In this paper we do not include low-energy photons. This is because \cite{Reissl2018} showed that infrared photons are unable to couple to the ISM, and thus can not strongly influence the evolution of \hii regions. The reason is that photons shift to lower energies as they are absorbed and remitted at longer wavelengths by dust, decreasing their momentum transfer rate over time in all but the most massive clouds. \cite{Rahner2017} further considered a full range of feedback forces and found that when properly accounting for opacity in dynamic models, IR radiation pressure was at most a small perturbation.

\subsection{Interacting Processes}
\label{introduction:interactingprocesses}

Various work has been done to describe the interaction between winds and photoionised \hii regions. \cite{Dyson1977} used adiabatic wind solutions to trace the expansion of blister \hii regions with embedded winds, while \cite{Capriotti2001} produced more detailed solutions for winds and a photoionised \hii region in a uniform external medium, finding that winds are very rarely the principal drivers of the expansion. \cite{Krumholz2009} performed an analysis of various modes of feedback by comparing them to the \cite{SpitzerLyman1978} solution for a D-type photoionised \hii region in a uniform medium, while \cite{Haworth2015} did the same for the detailed microphysics driving D-type \hii regions.

\cite{Haid2018} found good agreement between simple analytic models and simulations of \hii regions and stellar winds in uniform environments. In their paper, as well as the model of \cite{Krumholz2009} and the simulations of \cite{Dale2014}, photoionisation is the most important process to capture in most dense cloud environments studied. Outside the cloud, however, they argue that photoionisation should have a minimal effect since background radiation will heat the diffuse gas to temperatures near the equilibrium temperature of photoionised gas.

A simple approach previously employed has been to add the pressures from each dynamical process driving feedback bubbles as if they acted independently from the other forces, for example by \cite{Murray2010}. This is a reasonable assumption if all but one process is negligible at a given time, but breaks down if complex interactions are considered. For example, as in \cite{Pellegrini2007}, while radiation pressure adds a force that is propagated to the edge of the \hii region, the departure from a constant density \hii region increases the recombination rate of the \hii region, shrinking the radius of the region as a result. \rev{Since a full physical model has no linear algebraic solution, various numerical analytic solutions exist that capture the various physics involved \citep[e.g.][]{Pellegrini2009,Yeh2012,Verdolini2013,Rahner2017}.}

We conclude that \hii regions are complicated phenomena that require careful consideration in order to avoid oversimplifying assumptions about their structure and dynamics. In this paper we focus specifically on how winds and radiation pressure shape photoionised \hii regions expanding into self-gravitating power-law gas density fields, treating each process as perturbations to our dynamical equations. We calculate where such perturbations should become large enough to demand more complicated numerical solutions.

\subsection{Paper Outline}

We start by briefly outlining the conceptual structure of an \hii region in Section \ref{overview}. We then give equations used to describe the cloud into which the \hii region expands in Section \ref{cloud}. In Section \ref{photoionisation_only}, we introduce a set of formulae to describe the expansion of ionisation fronts driven purely by photoionisation. We then derive corrections to these expressions to account for the role of winds (Section \ref{winds_in_uv}), radiation pressure (Section \ref{radiation_pressure}), photon breakout and gravity (Section \ref{photon-breakout-and-gravity}). For each additional effect we calculate a coefficient that characterises how important the effect is compared to a solution without that effect. We then compute solutions to the problem using physically motivated cloud properties and stellar evolution models. In Section \ref{model} we describe how we set up these models, and in Section \ref{results} we summarise the results of these solutions. Finally, we discuss the \rev{context and} significance of our findings in Section \ref{discussion}.

\section{A Typical \hii Region}
\label{overview}

\begin{figure}
	\centerline{\includegraphics[width=\columnwidth]{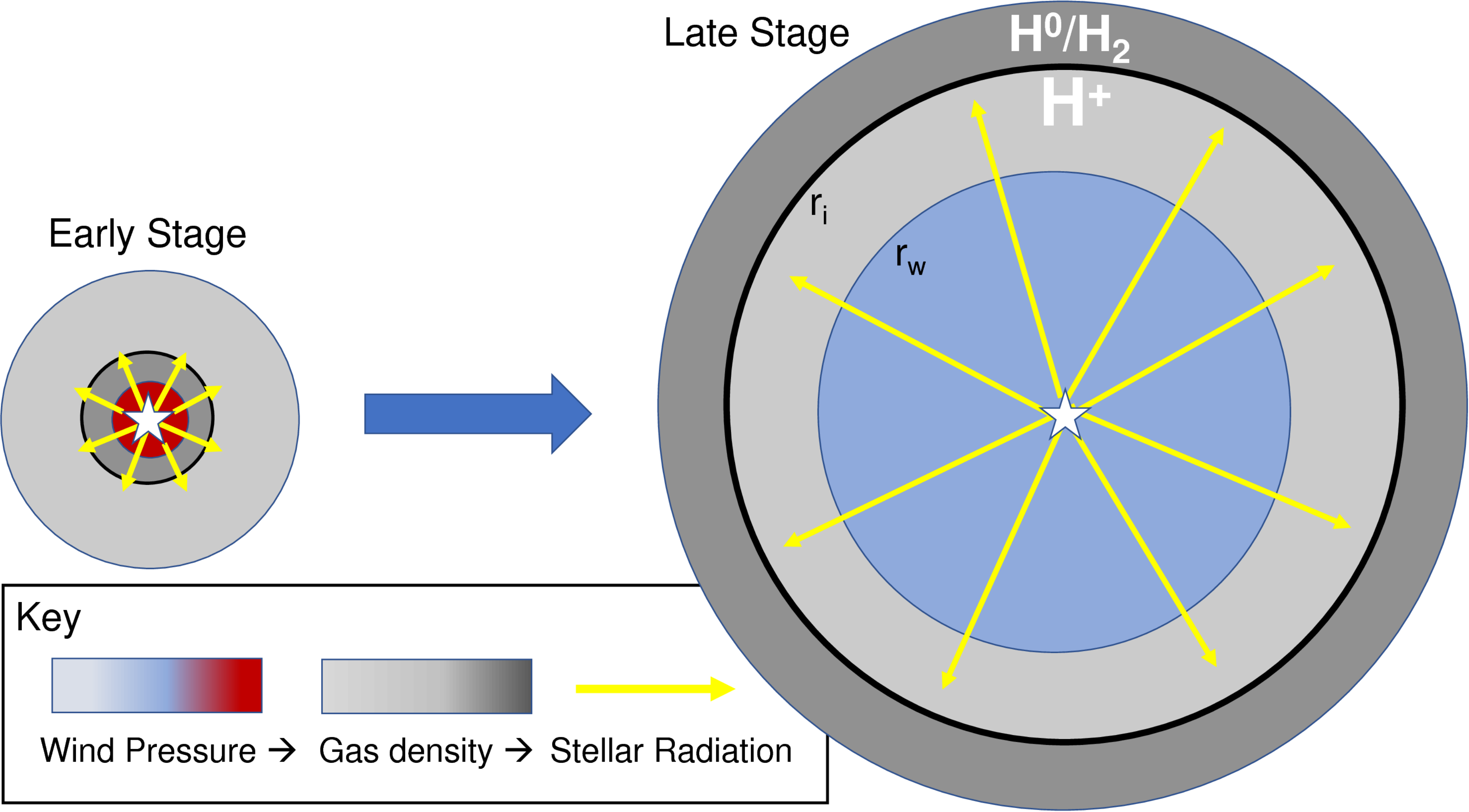}}
	\caption{Schematic showing the components of the modelled \hii region. Figure adapted with permission from \protect\cite{Rahner2017}.}
	\label{fig:schematic}
\end{figure}

\begin{figure}
	\centerline{\includegraphics[width=\columnwidth]{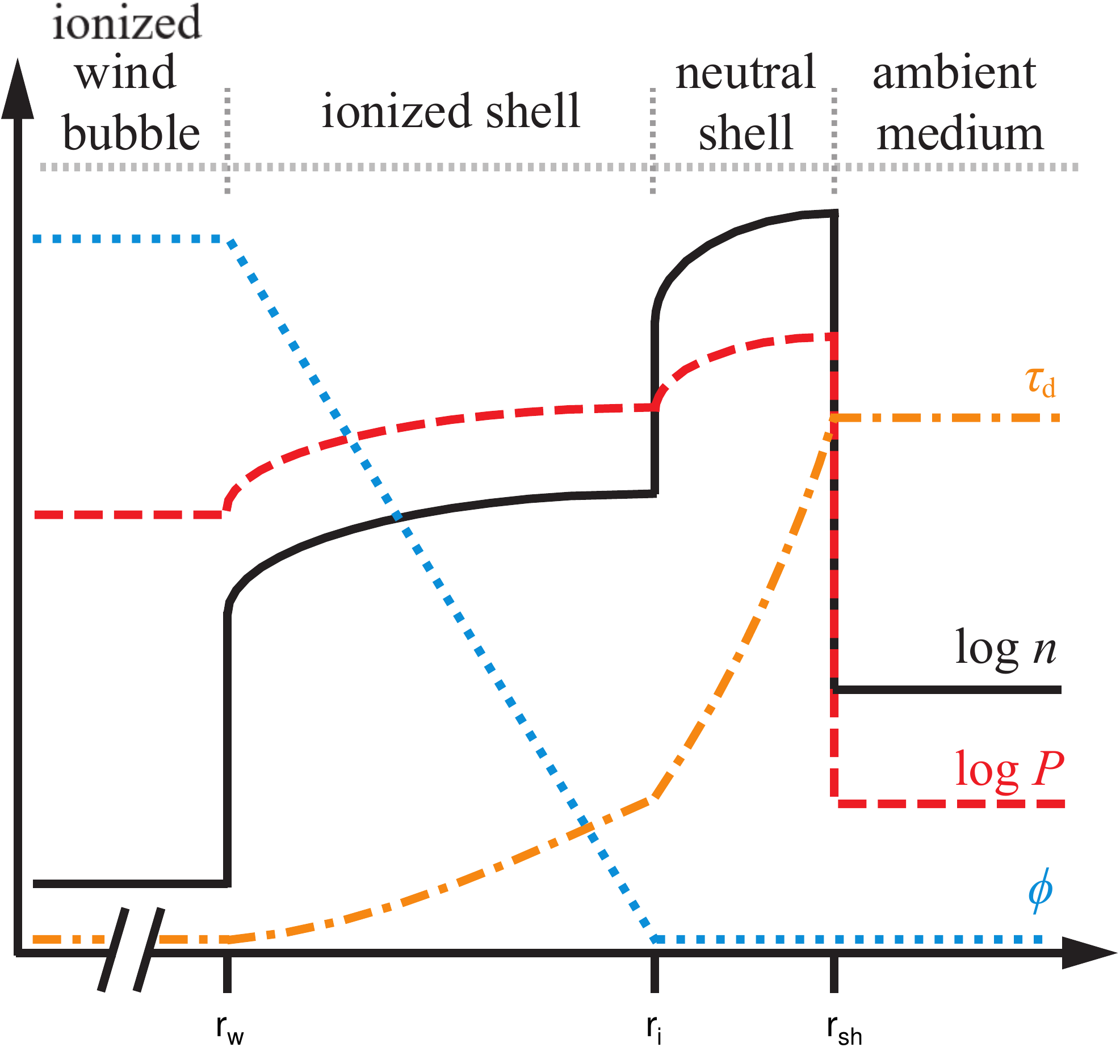}}
	\caption{Schematic showing the typical density ($n$), pressure ($P$), ionising photon flux ($\phi$) and optical depth of the dust ($\tau_d$) profiles in the \hii region and outside. Figure adapted with permission from \protect\cite{Rahner2017}.}
	\label{fig:profiles}
\end{figure}

In this Section we summarise the components of an \hii region to illustrate how our model is constructed. We do not provide details or justifications for model choices in this Section, but link to the parts of the text where each effect is discussed in more detail.

The structure of an \hii region is illustrated in Figure \ref{fig:schematic}. Typical radial distributions of density $n$, pressure $P$, photon flux $\phi$ and optical depth $\tau_d$ are given in Figure \ref{fig:profiles}. The radial components of this profile are as follows:

\begin{enumerate}

\item At $r=0, t=0$, the OB star or cluster of stars forms at a density peak in a volume of dense, neutral cloud material (Section \ref{cloud}).

\item Stellar winds have the same temperature as the surface of the star, reaching terminal velocities of 1000 to 3000 km/s (see Section \ref{model:stellarevolution}). This material shocks against the ambient medium, creating a bubble of hot, diffuse, collisionally ionised gas out to radius $r_w$ (Section \ref{winds_in_uv}).

\item Ionising UV photons are emitted from the star. They pass through the collisionally-ionised wind bubble and photoionise a volume of gas from $r_w$ to $r_i$. A typical assumption is that these regions have \rev{a} constant density (Section \ref{photoionisation_only}). Radiation pressure introduces a density gradient in the \hii region, which in certain regimes cannot be ignored \rev{(see Section \ref{radiation_pressure:strong} for a discussion)}.

\item A dense, neutral shell is swept up between $r_i$ and $r_{sh}$, beyond which lies the unshocked cloud material. If sufficient photons are emitted, this shell can become fully ionised, causing photons to break out of the \hii region and stream into the external medium (Section \ref{photon-breakout}). $r_{sh}$ can also shrink via self-gravity or accretion from the external medium (Section \ref{gravity}). In this paper we assume that $r_{sh} \simeq r_i$.

\end{enumerate}

\section{Host Environment of the \hii Region}
\label{cloud}

In this Section we discuss the density distribution of the initially neutral gas into which the \hii region expands.

\subsection{A Power-Law Density Profile}

The majority of massive stars are thought to form in molecular clouds. These objects have been widely studied in the literature to date \citep[see reviews by][]{MacLow2004,Ballesteros-Paredes2007,Hennebelle2012,Klessen2016}. They are often modelled as fractal structures based on analysis of observed clouds \citep{Cartwright2004,Hennebelle2012,Li2015,Jaffa2018}. \cite{Larson1981} argues for a hierarchy of overdensities with defined power-law relations in size, mass, density and velocity dispersion. These overdensities collapse under gravity, and form peaked density profiles inside which protostars form.

In the recent simulations of \cite{Lee2018}, star-forming cores are found to have a roughly power-law density profile $n \propto r^{-w}$ where $w \simeq 2$, otherwise called a \SIS. Observational studies by \cite{Kauffmann2010a} find a similar trend, based on inference from 2D column density maps. This profile has also been used in previous theoretical studies of \hii regions, e.g., by \cite{Shu2002} and \cite{Alvarez2006} among others. 

In this Section and Section \ref{photoionisation_only} we give solutions for an arbitrary $w$. In order to provide tractable solutions, Sections after that all assume $w=2$. We ignore the role of accretion in changing the form of this profile, and assume that once the \hii region forms its expansion is rapid enough to neglect additional mass from outside accretion. Future use of these models should explore this issue.

We define the density profile of the cloud
\begin{equation}
n(r) = n_0 (r / r_0)^{-w}
\label{cloud:profile}
\end{equation}
where $n(r)$ is the number density of the cloud at radius $r$ and time $t$, $w$ is the exponent of the power law, $n_0$ and $r_0$ are a characteristic number density and radius. The quantities $r_0^w$ and $n_0$ are degenerate. In this paper we use $r_0$ as some arbitrary boundary at which we measure the cloud properties, and so $n_0 = n(r_0)$.

The mass density of the cloud is given by 
\begin{equation}
\rho(r) = n(r) \frac{m_H}{X},
\end{equation}
where $m_H$ is the mass of a hydrogen atom and $X$ is the hydrogen mass fraction. The mass of the cloud (ignoring any central stellar mass) integrated out to a radius $r$ is then
\begin{equation}
M(<r) = \frac{4 \pi}{3 - w} n_0 r_0^w r^{3-w} \frac{m_H}{X}.
\label{cloud:mass}
\end{equation}

\cite{Lombardi2010} argue that structures inside observed nearby molecular clouds are defined by mass contained within a given cut-off surface density $\Sigma(r)$. At radius $r$ and for $w=2$, $\Sigma(r)$ is given by:
\begin{equation}
\Sigma(r) = \pi n_0 r_0^2 r^{-1} \frac{m_H}{X}.
\label{cloud:surfdens} 
\end{equation}
\citep[see][]{Binney2008}. Defining $\Sigma_0 \equiv \Sigma(r_0)$ and $M_0 \equiv M(<r_0)$, we can write for $w=2$:
\begin{equation}
n_0 r_0^2 \frac{m_H}{X} = \frac{1}{2 \pi} (M_0 \Sigma_0)^{1/2}.
\end{equation}
The mass inside radius $r$ then becomes
\begin{equation}
M(<r) = 2 (M_0 \Sigma_0)^{1/2} r.
\label{cloud:mass_isothermal}
\end{equation}

We can solve these equations to relate $r_0$ and $n_0$ to $\Sigma_0$ and $M_0$:
\begin{equation}
r_0 = \frac{1}{2} \sqrt{\frac{M_0}{\Sigma_0}},
\label{cloud:r0}
\end{equation} 
\begin{equation}
n_0 = \frac{2}{\pi} \frac{X}{m_H} M_0^{-1/2} \Sigma_0^{3/2}.
\label{cloud:n0}
\end{equation}

\section{Thermal Expansion of a Constant Density Photoionised \hii Region}
\label{photoionisation_only}

In this Section we introduce our model for the behaviour of photoionisation fronts in self-gravitating power-law density profiles without winds or radiation pressure. See Section \ref{introduction:hiiregions} for a discussion of previous analytic work on this topic.

In addition to the power-law structure of the neutral cloud in Equation \ref{cloud:profile}, we must invoke two additional conditions:
(1) that the \hii region contained within the photoionisation front is in photoionisation equilibrium, (2) and the thermal pressure of the photoionised gas is balanced by the ram pressure on the shell from the external medium. In Section \ref{photon-breakout} we discuss when these conditions break down.

Condition (1), photoionisation equilibrium, is satisfied by
\begin{equation}
\frac{4 \pi}{3} r_i^3 n_i^2 \alpha_B = Q_H
\label{photo:equilibrium}
\end{equation}
as discussed by \cite{Stromgren1939}, where $r_i$ is the radius of the photoionisation front, $n_i$ is the number density in the photoionised gas, $\alpha_B$ is the recombination rate of the photoionised gas (= 2 to $3\times10^{-13}$ cm$^{3}$/s in typical solar conditions), and $Q_H$ is the emission rate of ionising photons. In this Section, as in \cite{SpitzerLyman1978} and others, we assume that the density in the photoionised gas is constant due to thermal mixing. \rev{In these calculations we typically assume that the ions are well represented by singly ionised hydrogen. We ignore a small correction in the electron number density due to multiple ionisation of helium, metal species and dust. In this Section we ignore the effects of radiation pressure, which we discuss in Section \ref{radiation_pressure}.}

Condition (2), pressure equilibrium, is given by
\begin{equation}
n_i c_i^2 = n(r_i) (\dot{r}_i + v_0)^2 ,
\label{photo:externalpressure}
\end{equation}
where $c_i$ is the \rev{isothermal} sound speed in the photoionised gas (on the order of 10 km/s for gas at $10^4$ K), $n(r_i)$ is a power law defined as in Equation \ref{cloud:profile}, and $v_0$ is a generic term used to describe pressure forces from gravity and accretion flows from the external medium onto the shell as in \cite{Raga2012}. We assume that all velocities are in the radial direction and so values of $v_0$ are negative, representing either a resistive force on the expansion or accretion onto the shell. In Appendix \ref{appendix:pressure_velocity} we derive forms for this $v_0$, which is independent of $r$. 

We define a stall radius $r_{stall}$ where $\dot{r_i} = 0$, and introduce a scale-free radius $R \equiv r_i / r_{stall}$ (and $\dot{R} \equiv \dot{r}_i / r_{stall}$). Solving these equations (see Appendix \ref{appendix:stalled_photoionisation}) gives
\begin{equation}
\dot{r}_i = v_0 (R^{\frac{2 w - 3}{4}} - 1).
\label{photo:dynamical_equation}
\end{equation}
In a uniform medium ($w=0$), this equation has a \textit{stable} point at $r_{stall}$, where $R$ will tend to 1 over time. \rev{Note that $v_0$ is of the order of the escape velocity $\sqrt{2 G M/R}$. Using Equation \ref{cloud:mass}, this gives $v_0 \propto R^{1-w/2}$. $v_0$ is thus constant with $R$ where $w=2$.}

\begin{figure}
	\centerline{\includegraphics[width=0.95\columnwidth]{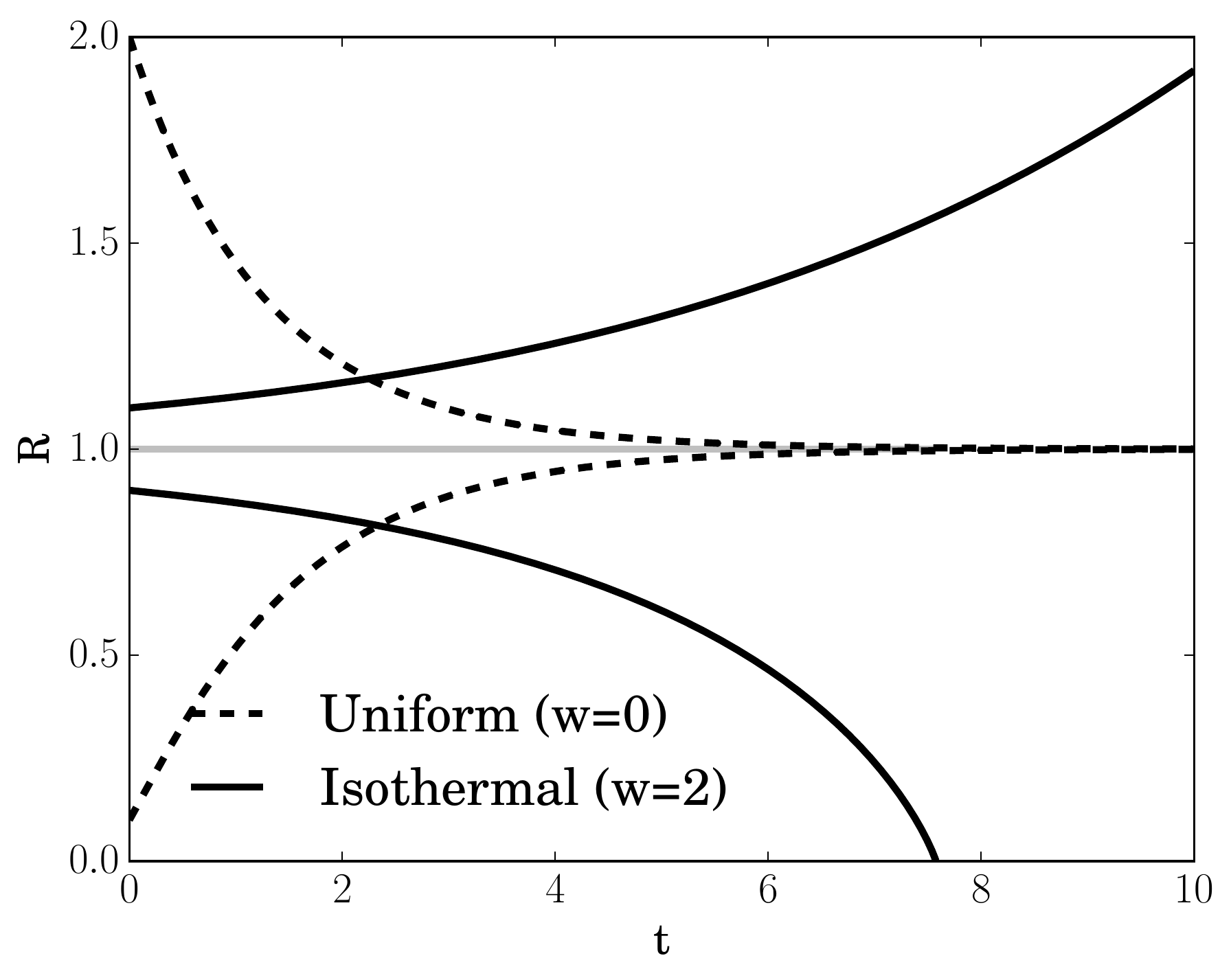}}
	\caption{Solutions to Equation \protect\ref{photo:dynamical_equation} for the expansion of an \hii region with a ram pressure term based on gravity or accretion \rev{using scale-free units}. Here, $r_{stall} = r_{launch}$ = 1 (grey solid line) \rev{and $v_0 = R^{1-w/2}$}. This figure demonstrates that in the case of a uniform gas field, $r_{stall}$ is a \textit{stable} point which solutions converge to. For steeper potentials, the solution tends to zero or infinity, i.e. $r_{stall}$ is an \textit{unstable} point. $r_{stall}$ is thus no longer a true ``stall'' point, but instead a launching radius $r_{launch}$.}
	\label{fig:stallplot}
\end{figure}

For $w>3/2$, the exponent $(2 w - 3)/4$ in Equation \ref{photo:dynamical_equation} becomes positive. Under these conditions, $r_{stall}$ becomes an \textit{unstable} equilibrium point in a \textit{bistable} system with attractors at $R = 0$ and infinity. \textit{In other words, the ionisation front will either accelerate forever or be crushed depending on the initial radius of the photoionisation front}. We illustrate this behaviour in Figure \ref{fig:stallplot}. 

Under these conditions, $r_{stall}$ is no longer a true ``stall'' point, but instead a ``launching'' radius, $r_{launch} \equiv r_{stall}$. In other words, if $r > r_{launch}$, the solution will accelerate over time. We use this terminology for the rest of the paper, since we focus largely on profiles where $w=2$.

In Appendix \ref{appendix:stalled_photoionisation}, we derive Equation \ref{photo:dynamical_equation} in terms of $t_{ff,0}$, which is the freefall time at density $n_0$ \rev{where $w=2$}:
\begin{equation}
\dot{R} = R_0 \frac{\pi}{2} \frac{\sqrt{3f}}{t_{ff,0}}(R^{1/4} - 1)
\end{equation}
where $R_0 \equiv r_0 / r_{launch}$ and $f \equiv (v_0/v_{esc})^2$.

The free fall time $t_{ff,0}$ is thus a significant factor in setting the timescale for the expansion of the \hii region \citep[see also][]{Geen2016}.

%In the next Section we add winds to this picture in order to determine how they modify the dynamics of these systems.

\section{Winds in UV Photoionised Bubbles}
\label{winds_in_uv}

Stellar winds are outflows of material from stars driven by radiation from inside the star. Winds from massive stars reach a terminal velocity of 1-4 times the escape velocity at the surface of the star \citep[e.g.][]{Vink2011}. This velocity can be well over 1000 km/s. 

The winds initially free-stream but at some radius shock against the ambient medium. The temperature of the shocked gas is typically above $10^6$ K, and the temperature of the thermalised gas is in pressure equilibrium with the speed of the free-streaming phase \citep[see, e.g.][for a discussion of this transition]{Weaver1977}. 

Since the temperature of this gas is typically well above the limit to be collisionally ionised, the UV photons from the star \rev{are not absorbed by the wind bubble}. The wind is thus always embedded inside a photoionised volume, although for strong winds and weak ionising sources, the photoionised region can be a thin shell around a wind-blown bubble. 

In this Section we first discuss how winds affect the structure of photoionised \hii regions. We then calculate the relative impact of winds and photoionisation on the dynamics of \hii regions. \rev{To provide tractable solutions, all further equations in this paper take $w=2$.}

\subsection{Pressure from Stellar Winds}

We have a stellar source with mass loss rate $\dot{m}_w$ and terminal velocity $v_w$. This gives us a wind luminosity $L_w \equiv \frac{1}{2}\dot{m}_w v_w^2$ and a wind momentum deposition rate $\dot{p}_w \equiv \dot{m}_w v_w$\footnote{Throughout this paper, small $p$ denotes a momentum term, and large $P$ a pressure term.}.

There are two models for the pressure exerted by stellar winds onto an external medium given by \cite{Silich2013}. In the first, energy-conserving model, minimal radiative losses are assumed as in \cite{Weaver1977}, and the bubble is supported by thermalised gas that stores the energy emitted by the stellar source over its lifetime,
\begin{equation}
P_{w,e} = \left( \frac{k_w^2 L_w^2 \rho_i}{r_w^4} \right)^{1/3} ,
\label{wind:pressure:energy}
\end{equation}
where $r_w$ is the radius of the wind bubble, \rev{$\rho_i$ is the mass density in the ionised gas $n_i m_H / X$,} and $k_w$ is a constant dependent on the adiabatic index $\gamma$ \rev{in the wind bubble},
\begin{equation}
k_w^2 = \frac{7}{25} \left ( \frac{375 (\gamma - 1)}{28 \pi (9 \gamma - 4)} \right)^2.
\end{equation}

In the second, momentum-conserving model, the wind bubble cools strongly due to the high density of the gas. Storage of energy from previous wind emission is minimal. The outward force instead comes directly from the momentum injection rate of the star. This model is given by 
\begin{equation}
P_{w,m} = \frac{\dot{p}_w}{4 \pi r_w^2}.
\label{wind:pressure:momentum}
\end{equation}

The precise pressure from the wind bubble will lie somewhere between these two solutions \citep[see, e.g.][]{Silich2018}. If the cooling time is much longer than the lifetime of the stellar source, the energy-conserving solution will apply. \cite{MacLow1988} argue that this condition applies for galactic superbubbles in more diffuse media. For much denser environments, where the cooling time is much shorter than the lifetime of the stellar source, the momentum-conserving solution is more applicable. The cooling timescale for the wind bubble is given in \cite{MacLow1988} as
\begin{equation}
t_c = (2.3 \times 10^4 \mathrm{yr}) n_i^{-0.71} (L_w / 10^{38} \mathrm{erg/s})^{0.29}
\label{wind:cooling_time}
\end{equation}
assuming that the wind bubble is expanding into the photoionised region with a density of $n_i$ \rev{and the main cooling mode is mixing with the background material}.

To get an estimate of how much $n_i$ is reduced with respect to the external density field with a $1/r^2$ profile, we can use Equations \ref{cloud:profile} and \ref{photo:equilibrium} to write:
\begin{equation}
\frac{n_i}{n(r_i)} = \left( \frac{3 Q_H r_i}{4 \pi \alpha_B} \right)^{1/2} \frac{1}{n_0 r_0^2}.
\label{wind:density_contrast}
\end{equation}
If we take $r_i = r_0$, this ratio is above 0.1 for typical values used in this work. The internal density of the \hii region is thus not significantly reduced compared to the external density. This is because the \hii region expands into a cloud with a density profile that decreases with radius, and so the photoionised gas remains at a high density compared to its surroundings. In a uniform region, where the external density does not change, the \cite{SpitzerLyman1978} solution leads to much lower values of $n_i / n(r_i)$ over time. See Section \ref{results:evolution-of-an-example-hii-region} for an example of the evolution of $n_i$ around a single massive star.

The values of $t_c$ at molecular cloud densities is typically short, on the order of 1 to 10 kyr. We expect the momentum-driven pressure in Equation \ref{wind:pressure:momentum} to be more representative under the conditions considered in this paper, and use this form from now on. An algebraic model with fully self-consistent radiative cooling is difficult to obtain. \cite{Rahner2019} recently calculated a semi-analytic solution that treats the out-of-photoionisation-equilibrium radiative cooling in \hii regions with wind bubbles, which accurately models the transition from adiabatic to radiatively cooled winds. This is likely to be most important in the early stage of the \hii region evolution or in more diffuse regions where the density is lower, as in \cite{MacLow1988}. \cite{Rahner2019} also treat the long-term evolution of \hii regions after the first supernovae have occurred, which we omit in this work.

\subsection{Evolution inside \hii Regions}

Since the sound speed in the wind bubble is much higher than the sound speed in the photoionised gas ($> 1000$ km/s versus $\sim10$ km/s), we assume that $r_w$ reaches its equilibrium state quickly compared to the expansion rate of the photoionisation bubble. The radius of the wind bubble $r_w$ is thus always balanced by hydrostatic equilibrium between the wind pressure and the thermal pressure of the photoionised gas.

We modify Equation \ref{photo:equilibrium} to account for the collisionally-ionised wind bubble inside the \hii region,
\begin{equation}
\frac{4 \pi}{3} n_i^2 (r_i^3 - r_w^3) \alpha_B = Q_H.
\label{wind:photoequilibrium}
\end{equation}

The pressure condition on the external medium in Equation \ref{photo:externalpressure} remains valid when a wind bubble is included, since the wind bubble is contained entirely within the photoionised \hii region. The pressure from the wind in Equation \ref{wind:pressure:momentum} (or Equation \ref{wind:pressure:energy} in the energy-conserving case) balances the pressure in the photoionised gas as
\begin{equation}
P_w = n_i c_i^2 \frac{m_H}{X}.
\label{wind:windpressurebalance}
\end{equation}

In the momentum-conserving phase, this gives
\begin{equation}
\frac{\dot{p}_w}{4 \pi r_w^2} = n_i c_i^2 \frac{m_H}{X}.
\label{wind:windpressurebalance_momcons}
\end{equation}

We can substitute these equations (see Appendix \ref{appendix:wind}) as
\begin{equation}
(\dot{r_i} + v_0)^4 = A_w (\dot{r_i} + v_0) + A_i r_i,
\label{wind:dynamics}
\end{equation}
where $A_w$ and $A_i$ are constants relating to the wind and photoionisation respectively:
\begin{equation}
A_w = \left ( \frac{\dot{p}_w}{4 \pi} \frac{X}{m_H} \right )^{3/2} \left ( n_0 r_0^2 \right )^{-3/2} 
\label{wind:Aw}
\end{equation}
and
\begin{equation}
A_i = \frac{Q_H}{\alpha_B} \frac{3}{4 \pi}\left (c_i^2 \right )^{2}  \left ( n_0 r_0^2 \right )^{-2}  .
\label{wind:Ai}
\end{equation}

In Appendix \ref{appendix:pressure_velocity} we show that for a $1/r^2$ density profile, $v_0$ is constant. This means that the expansion rate $r_i$ of the \hii region is reduced by a constant factor $v_0$ at all times.

\subsection{How Important are Winds Dynamically?}

When the contribution from winds is equal to the contribution from photoionisation, we can write Equation \ref{wind:dynamics} as
\begin{equation}
(\dot{r_i}_{eq} + v_0)^4 = 2  A_i r_i = 2 A_w (\dot{r_i}_{eq} + v_0) ,
\label{wind:equalcontribution}
\end{equation}
where $\dot{r_i}_{eq}$ is the expansion rate under these conditions.

We can define a factor $C_w$ as the ratio of the contribution from winds and photoionisation. 
$C_w < 1$ if photoionisation is the principal driver in Equation \ref{wind:dynamics}, and $C_w > 1$ if winds are the principal driver. If Equation \ref{wind:equalcontribution} applies, $C_w=1$. Around this limit,
\begin{equation}
\begin{split}
C_w \equiv \frac{A_w(\dot{r_i} + v_0)}{A_i r_i} \simeq \frac{A_w (2 A_i r_i)^{1/4}}{A_i r_i} \\
= 2^{1/4} A_w (A_i r_i)^{-3/4} .
\end{split}
\label{wind:comparison}
\end{equation}
This can be expanded using Equations \ref{wind:Aw} and \ref{wind:Ai} to the form
\begin{equation}
C_w = 2^{1/4} \left( \frac{\dot{p_w}}{4 \pi} \frac{X}{m_H} \frac{1}{c_i^2} \right)^{3/2} \left( \frac{Q_H}{\alpha_B} \frac{3}{4 \pi} r_i \right)^{-3/4}.
\label{wind:coefficient}
\end{equation}

Setting $r_i=r_0$, we can calculate the relative effect of winds versus photoionisation in driving the expansion of an \hii region by the time it reaches a characteristic radius $r_0$ enclosing a mass $M_0$ with bounding surface density $\Sigma_0$, using fiducial values for the terms in Equation \ref{wind:coefficient} as listed in Section \ref{model}:
\begin{equation}
\begin{split}
C_w = 0.00931
\left( \frac{\dot{p}_w}{10^{28}~\mathrm{g.cm/s^2}} \right)^{3/2} 
\left( \frac{Q_H}{10^{49}~\mathrm{s}^{-1}} \right)^{-3/4} \times \\ 
\left( \frac{M_0}{100~\mathrm{M}_{\odot}} \right)^{-3/8}
\left( \frac{\Sigma_0}{100~\mathrm{M}_{\odot}~\mathrm{pc}} \right)^{3/8} 
\left( \frac{c_i}{10~\mathrm{km/s}} \right)^{-3} .
\end{split}
\label{wind:condition}
\end{equation}

The relative impact of winds depends on the factor $\dot{p}_w^2 / Q_H$. In other words, for a given stellar evolution model, the relative impact of winds grows as stars are added (i.e. where both quantities increase linearly).

\subsection{Ratio of Radii of Wind and Photoionisation Bubbles}
\label{wind:ratioofbubbles}

Substituting Equations \ref{wind:windpressurebalance_momcons} and \ref{wind:coefficient} into Equation \ref{wind:photoequilibrium}, we can find a relationship for the ratio between the radius of the photoionisation bubble $r_i$ and the radius of the wind bubble $r_w$. Note that $r_i$ is the radius of the \hii region as a whole, since the ionising photons always pass through the collisionally-ionised wind bubble:

%\begin{equation}
%\left (\frac{r_i}{r_w}  \right )^{4} = \frac{3}{4 \pi} \frac{Q_H}{\alpha_B} \left ( \frac{c_i^2}{\gamma} %\frac{m_H}{X} \frac{8 \pi}{\dot{p}_w} \right )^2  r_i + \frac{r_i}{r_w}
%\label{wind:radiusratio}
%\end{equation}

\begin{equation}
\left (\frac{r_i}{r_w}  \right )^{4} = 2^{1/3} C_w^{-4/3} + \frac{r_i}{r_w} .
\label{wind:radiusratio}
\end{equation}

For large values of $C_w$, $r_w \rightarrow r_i$. At $C_w=1$, $r_w/r_i = 0.79$. At $C_w=0.1$, this drops to $r_w/r_i = 0.43$, while $C_w=10$ gives $r_w/r_i = 0.98$. A reasonably large wind bubble can therefore exist even if winds are not dynamically significant, while for cases where winds are the main driver of the \hii region, the wind bubble will occupy most of the volume of the \hii region. A large wind bubble is therefore not always an indicator that winds are dynamically important.

\subsection{Density of Photoionised Gas}
\label{wind:density-of-photoionised-gas-and-observational-significance}

By substituting the expression for $r_w$ derived from Equation \ref{wind:windpressurebalance_momcons} into Equation \ref{wind:photoequilibrium}, we can write the density of the photoionised gas as
\begin{equation}
n_i^2 r_i^3 = \left( \frac{\dot{p_w}}{4 \pi} \frac{X}{m_H} \frac{1}{c_i^2} \right)^{3/2} n_i^{1/2} + \frac{Q_H}{\alpha_B} \frac{3}{4 \pi}.
\label{wind:density}
\end{equation}
In the absence of stellar wind, this becomes Equation \ref{photo:equilibrium} describing photoionisation equilibrium where the photoionised gas fills the spherical volume enclosed by $r_i$ ($n_i \propto r_i^{-3/2}$). If, however, winds become the principal driver of the \hii region (i.e. the second term in the right-hand side of Equation \ref{wind:density} is neglected), we obtain $n_i \propto (4 \pi r_i^2)^{-1}$. In these conditions, $r_w \rightarrow r_i$ and the photoionised gas is confined to a thin shell at $r_i$.

\rev{We discuss the observational significance of the changes to the radial structure and density of the \hii region in Section \ref{discussion:obs}.}

% Removed to cite other work that discusses this in more detail
%This is significant for observational studies because it sets the behaviour of observable quantities defining the emission from photoionised gas. For example, the ionisation parameter ${\cal U}$ which traces emission line diagnostics in photoionised gas, such as H$\alpha$, [O$_{II}]$, [O$_{III}$], [N$_{II}$], [S$_{II}$] and others, is often defined as
%\begin{equation}
%{\cal U} = \frac{Q_H}{4 \pi r_i^2 n_i c}.
%\end{equation}
%This condition scales best in the limit where $C_w \gg 1$, i.e. $n_i \propto r_i^{-2}$, and behaves more variably in photoionisation-driven \hii regions.

\section{Radiation Pressure}
\label{radiation_pressure}

The role of radiation pressure from ionising photons is different from stellar winds. Rather than acting on the inner surface of the photoionised region, radiation pressure distributes itself throughout the photoionised gas. In this section we calculate the relative effect of adding a radiation pressure term to the solutions derived in the paper so far. We treat radiation pressure as a perturbation to the constant-density \hii region model and discuss what happens when this perturbation grows to non-negligible levels.

\subsection{Relative importance of Radiation Pressure}
\label{radiation_pressure:relative_importance}

\rev{We introduce a quantity $C_{rp}$ to track the effect of radiation pressure as a perturbation to the photoionisation-driven solution. We use a similar method to \cite{Krumholz2009}, who consider the addition that radiation pressure makes to the total pressure from the \hii region at $r_i$. \cite{Krumholz2009} write the total additional pressure from radiation acting on the \hii region at $r_i$ as}
\begin{equation}
\Delta P_{rp} = \frac{f_{trap} L}{4 \pi r_i^2 c},
\label{radiation_pressure:luminosityeqn}
\end{equation}
\rev{where $L$ is the total bolometric luminosity of the star and $f_{trap}$ is a boosting factor to account for multiple scattering of photons. For significant losses of photons, $f_{trap}$ can be less than 1. We calculate whether the escape of ionising photons occurs in Section \ref{photon-breakout}.}

\rev{$L$ can be written in terms of the ionising photon output as}
\begin{equation}
L = \psi \langle h \nu \rangle Q_H,
\end{equation}
\rev{where $\langle h \nu \rangle$ is the average energy of photons emitted by the star and $\psi$ is a factor that accounts for the fraction of the spectrum below 13.6 eV. \cite{Krumholz2009} adopt $\psi=1$ in the case of output from massive stars, which emit significant quantities of ionising radiation.}

\rev{Setting $P_0$ to the thermal pressure in the \hii region from photoionisation using Equation \ref{wind:windpressurebalance}, we can write}
\begin{equation}
C_{rp} \equiv \frac{\Delta P_{rp}}{P_0} = \frac{Q_H}{4 \pi r_i^2} \frac{f_{trap} \langle h \nu \rangle}{c} \left (c_i^2 \frac{m_H}{X}  \right )^{-1}\frac{1}{n_{i}}.
\label{radiation_pressure:crpfull}
\end{equation}

\rev{In the absence of winds, this becomes}
\begin{equation}
C_{rp,nowind} = \left ( \frac{ Q_H}{12 \pi \alpha_B \phi_d} \right )^{1/2} \frac{f_{trap} \langle h \nu \rangle}{c} \left(c_i^2 \frac{m_H}{X} . \right )^{-1} r_i^{-1/2}
\label{radiation_pressure:crpsimple}
\end{equation}
\rev{A term $\phi_d$ is introduced to account for losses to photoionised gas pressure from dust. This is equivalent to equation 7 in \cite{Krumholz2009}, who compute $C_{rp}$ in terms of $n_i$ rather than $r_i$. We adopt a fiducial $f_{trap}=2$ and $\phi_d=0.73$ from \cite{Krumholz2009} (a more accurate estimate of these quantities requires a numerical solution to the full radiative transfer equations). Equation \ref{radiation_pressure:crpsimple} can thus be written as}
\begin{equation}
\begin{split}
C_{rp,nowind} = 0.047
\left( \frac{\langle h \nu \rangle}{13.6 eV} \right) 
\left( \frac{c_i}{10~\mathrm{km/s}} \right)^{-2} 
\left( \frac{Q_H}{10^{49}~\mathrm{s}^{-1}} \right)^{1/2} \\
\left( \frac{r_i}{1 \mathrm{pc}} \right)^{-1/2}.
\end{split}
\label{radiation_pressure:condition}
\end{equation}
\rev{As in \cite{Krumholz2009}, radiation pressure is most important in small \hii regions with high luminosity sources.}

\rev{In Section \ref{results}, we solve for $C_{rp}$ using Equation \ref{radiation_pressure:crpfull} at a given $r_i$ and $n_i$. We use Equation \ref{wind:windpressurebalance_momcons} to calculate $n_i$, which requires Equation \ref{wind:radiusratio} to calculate $r_w$. Equation \ref{wind:radiusratio} does not have a simple algebraic solution, so we rely on a numerical solution to this particular equation. We discuss the effects of the inclusion of a wind bubble on the calculation of $C_{rp}$ in Section \ref{results:evolution-of-an-example-hii-region}.}

\subsection{\hii regions with strong radiation pressure}
\label{radiation_pressure:strong}

\rev{As the relative strength of radiation pressure increases, the structure of the \hii region changes. For increasing radius $r$ inside $r_i$, radiation pressure accumulates. This must in turn be balanced by increasing density as $r \rightarrow r_i$. For cases with very strong radiation pressure, as in small or very dense regions, most of the material in the \hii region is pushed towards a thin shell at $r_i$. This in turn allows the wind bubble to expand outwards. The combination of these equations is complex and requires a numerical solution \citep{Pellegrini2007,Pellegrini2009}.}

\rev{A parameter space exploration using numerical solutions for \hii regions with radiation pressure is given in \cite{Draine2011}. They do not include stellar winds but do include dust and non-ionising radiation. In Figure 5 of their paper, they give parameters for where the RMS density of the \hii region diverges significantly from the mean density (i.e. the \hii region can no longer be considered uniform density). Such density gradients are also visible in papers that compute numerical solutions including winds, magnetic fields, self-gravity and other effects \citep[e.g.][]{Pellegrini2007,Yeh2012,Rahner2019}.}

\rev{For values where $C_{rp}$ approaches 1, we expect winds and radiation pressure to move the \hii region's structure and evolution away from a simple algebraic solution. In this regime, a numerical solution is required. In Section \ref{results} we give the parameter space where this is likely to happen.}

\rev{The precise values of $f_{trap}$, $\phi_d$ and $\psi$ affect how effective radiation pressure is. This is not a trivial thing to calculate. In this work we pick fiducial values from \cite{Krumholz2009}, but there is still debate in the literature about how efficiently radiation pressure from photons is coupled to the gas. The simulations of \cite{Krumholz2012c} and \cite{Krumholz2013}, for example, show that instabilities generated in the gas and dust create cavities that allow radiation to escape more efficiently, reducing $f_{trap}$. \cite{Davis2014} and \cite{Rosdahl2015} confirm this with similar calculations using different radiation tracing methods. These effects are particularly important when considering the contribution of infrared photons. By contrast, the simulations of \cite{Kimm2019} show that Lyman alpha (Ly$\alpha$) photons can be generated through recombination of other ionising photons and scatter very efficiently, which in principle would boost $f_{trap}$. Photoelectric heating of dust grains has also been proposed as a feedback mechanism \citep{Forbes2016}. Radiation is thus a complex subject that requires careful modelling.}

\rev{In the next Section we discuss the role of gravity and photon breakout on the \hii region.}

\section{Photon Breakout and Gravity}
\label{photon-breakout-and-gravity}

In this Section we review additional terms presented in other work that affect the ionisation front. Firstly, we discuss the inability for the D-type solution to contain the ionising photons leading to rapid photoionisation of the whole clump/cloud. Secondly, we discuss the cases where gravity and accretion are able to stop the expansion of the \hii region.

\subsection{Photon Breakout}
\label{photon-breakout}

\cite{Franco1990} argue that for power-law external density fields where $w>3/2$, runaway photoionisation occurs, leading to a ``champagne'' flow that rapidly photoionises the entire cloud. If some process such as stellar winds create an initial dense shell, a D-type \hii region can be contained even in steeper density fields than this. However, once the champagne flow phase begins, the ionisation front expands in the R-type mode \citep{KahnF.D.1954}, which is solved in the relativistic limit by \cite{Shapiro2006}. After this point, the whole cloud is overtaken by the ionisation front and expands thermally at a few times the speed of sound in the photoionised gas \citep[see equation 24 in][]{Franco1990}. 

\cite{Alvarez2006} find the limit where this occurs, assuming the internal structure of the \hii region derived by \cite{Shu2002} in a $w=2$ cloud (we assume a similar density field for most of this paper). We can rewrite Equation 5 in \cite{Alvarez2006} for the radius at which the ionisation front breaks out of the D-type phase and enters a champagne phase $r_B$, in terms of $M_0$ and $\Sigma_0$ as
\begin{equation}
\begin{split}
r_B = 0.75~\mathrm{pc} 
\left( \frac{M_0}{100~\mathrm{M}_{\odot}} \right) 
\left( \frac{\Sigma_0}{100~\mathrm{M}_{\odot}~\mathrm{pc}^{-2}} \right) \\
\left( \frac{Q_H}{10^{49}~\mathrm{s}^{-1}} \right)^{-1}
\left( \frac{c_i}{10~\mathrm{km/s}} \right)^{4} .
\end{split}
\end{equation}

Comparing this to the radius of the cloud $r_0$ (see Equation \ref{cloud:r0}) and derive a ratio $C_{B} \equiv r_0 / r_B$,
\begin{equation}
\begin{split}
C_B = 0.8 
\left( \frac{M_0}{100~\mathrm{M}_{\odot}} \right)^{-1/2} 
\left( \frac{\Sigma_0}{100~\mathrm{M}_{\odot}~\mathrm{pc}^{-2}} \right)^{-3/2} \\
\left( \frac{Q_H}{10^{49}~\mathrm{s}^{-1}} \right)
\left( \frac{c_i}{10~\mathrm{km/s}} \right)^{-4} 
\end{split}
\label{breakout:condition}
\end{equation}
where $C_B > 1$ means that a breakout occurs before the ionisation front reaches $r_0$, assuming a \cite{Shu2002} solution for the photoionised gas. This is possible in small clouds with low surface densities and strong ionising photon sources with a low ionising sound speed. We give some physical scenarios for this in Section \ref{results}. The steep power-law dependence on $c_i$ means that efficient cooling found at super-solar metallicities may also allow a breakout.

We do not provide an algebraic form for the photon escape fraction from the \hii region in the presence of winds, radiation pressure and radiative cooling, since this requires complex numerical solutions such as those given by \cite{Rahner2017}. Instead, we offer the following comments:
\begin{enumerate}
	\item Photon leakage depends on the integration of $n(r,t)^2$ out to the ionisation front, including the shell around the ionisation front. This is governed by radiative cooling out of photoionisation equilibrium (as well as any non-spherical geometry, ignored here). Since winds drive the early expansion of the \hii region (c.f. Equation \ref{wind:dynamics}), a dense shell can already have formed, which will more effectively trap the \hii region.
	\item Photon leakage does not immediately cause photoionisation to become ineffective, since the wind must still expand into a dense, albeit photoionised cloud. In both cases the ionisation front expands rapidly (faster than $c_i$).
	\item Once the entire cloud has been ionised and $r_i$ enters the warm interstellar medium (WIM), there will be no well-defined \hii region boundary. Instead, only a wind-blown bubble may be observable as a distinct object, creating the misleading impression that the feedback from the star is driven solely by winds.
	\item We neglect here the role of fragmentation, which is included in the semi-analytic model of \cite{Rahner2019}.
\end{enumerate}

\subsection{The Role of Gravity and Accretion}
\label{gravity}

In this Section we discuss the role of resistive forces such as gravity and ram pressure from accretion on the shell around the ionisation front, which we have to this point characterised as a velocity term in the dynamical equations. In Appendix \ref{appendix:pressure_velocity} we derive forms for these terms.

In Section \ref{photoionisation_only}, we establish that in a $w=2$ density profile, the stall radius becomes a ``launching radius'' $r_{launch}$, above which $r_i$ accelerates over time. Rearranging Equation \ref{wind:dynamics} and setting $\dot{r}_i=0$ and $r_i = r_{launch}$ (for the sake of simplicity we neglect radiation pressure), we arrive at
\begin{equation}
r_{launch} = \frac{v_0 \left(v_0^3 - A_w\right)}{A_i} 
\label{wind:rstall}
\end{equation}

There is no good scale-free way to characterise $r_{launch}$ and determine whether the ionisation front will expand or be crushed, since this criterion depends on smaller scales governing evolution of the \hii region than we study here. We plot characteristic values for $r_{launch}$ in Section \ref{results} and discuss further what this means for the evolution of the \hii regions.

In Appendix \ref{appendix:gravityinthehiiregion} we calculate the ability for self-gravity inside the \hii region to set up a density gradient, similar to radiation pressure in Section \ref{radiation_pressure}. This effect is largely negligible except for in very dense, massive clouds with low temperature \hii regions.

%In Appendix \ref{appendix:rstall} we argue that if the escape velocity $v_{esc} > \frac{2}{\gamma} c_i$, the ionisation front must stall, unless winds or radiation pressure overcome the effect of gravity. This is similar to the argument of \cite{Dale2012} to within a factor of 2 (it depends also on the accretion rate, gravity of the central source, etc, which we ignore for now). Above this limit, accretion flows and gravity will crush the ionisation front, creating a flickering ultra-compact \hii region.
%
%In our cloud, the initial escape velocity is given by
%\begin{equation}
%v_{esc} = \sqrt{\frac{2 G M_0}{r_0}} = 0.656~\mathrm{km/s} \left( \frac{M_0}{100~\mathrm{M}_{\odot}} \right)^{1/4} \left( \frac{\Sigma_0}{100~ \mathrm{M}_{\odot}~\mathrm{pc}^{-2}} \right)^{1/4}
%\end{equation}

%We can compare this to $\frac{2}{\gamma} c$ as in Appendix \ref{appendix:rstall} to give a condition where gravity becomes a significant force in opposing the expansion of \hii regions
%\begin{equation}
%C_{esc} = 0.0656 \left ( \frac{c_i}{10 \mathrm{km/s}} \right )^{-1} \left( \frac{M_0}{100~\mathrm{M}_{\odot}} \right)^{1/4} \left( \frac{\Sigma_0}{100~ \mathrm{M}_{\odot}~\mathrm{pc}^{-2}} \right)^{1/4}
%\label{vesc:condition}
%\end{equation}
%
%For an isothermal cloud, it is thus difficult to find $v_{esc} > c_i \simeq 10$ km/s except for the densest and most massive clumps. For these denser clouds, the role of accretion flows and self-gravity will become important considerations in the expansion of the \hii region.

\section{Model Setup}
\label{model}

The parameter space invoked by the above set of equations is large. In this Section we model a physically-motivated subset of this parameter space to determine the importance of winds, radiation pressure and gravity in realistic conditions.

\subsection{Criteria for Complexity}
\label{model:conditions}

We focus on four important criteria governing the state of the \hii region: 
\begin{enumerate}
	\item $C_w>1$, the condition for winds to become more important than photoionisation in setting the expansion rate of the ionsiation front, given in Equation \ref{wind:condition},
	\item $C_{rp}>1$, the condition for the perturbation from radiation pressure to exceed the thermal pressure from photoionisation, given in Equation \ref{radiation_pressure:condition},
	\item $C_B>1$, the condition that ionising photons can break out of the shell around the \hii region, given in Equation \ref{breakout:condition}, and
	\item $r_{launch}$ compared to the initial radius of the \hii region, given in Equation \ref{wind:rstall}. This quantity is not scale-free and relies on gas dynamics and feedback physics on scales smaller than are studied in this paper. 
\end{enumerate}

In this Section we calculate each criterion for a range of typical values in solar and sub-solar metallicity environments. A fully coupled time-dependent solution requires an iterative numerical solution and is beyond the scope of this paper. We instead suggest in which context simple \hii region models considering only photoionisation feedback may be employed, and offer scope for future model exploration.

\subsection{Cloud Properties}
\label{model:cloudprops}

We focus on clouds with masses between $10^2$ and $10^5$ \Msolar for individual stars and binaries, and up to $10^7$ \Msolar for clusters. Smaller clouds are unlikely to form massive stars. We expect larger clouds to follow the trends indicated at the higher end of our mass range.

We compute each condition for a range of surface densities from $10^2$ to $10^4$ \Msolar/pc$^2$. The first value is typical of clouds in the solar neighbourhood \citep[e.g.][]{Lombardi2010}, while the latter value is typical of clouds in the Galactic Central Molecular Zone (CMZ) \citep[e.g.][]{Walker2016}. Note that the surface density threshold is also variable depending on which radius is selected, with higher densities closer to the peak of the density profile.

Where relevant, we assume that the cloud is not strongly accreting, i.e. $f = 1/2$ (see Appendix \ref{appendix:pressure_velocity}). In the case of large accretion rates, $f \rightarrow 3/2$, and $n(r)$ would become a function of time.

We invoke two models for selecting our sources of radiation and winds from stars. In the first case we use a source that is an individual star, or two stars of the same mass (to approximate an equal-mass binary). We then give results for these individual stars from 5 to 120 \Msolar. The motivation here is to study how \hii regions evolve around individual young massive stars or close binary systems.

In the second case, we use a source representing a cluster with an SFE of 10\% (i.e. the cluster mass is 10\% of the total gas mass inside $r_0$). To make this cluster, we draw randomly from a \cite{Chabrier2003} \IMF up to the cluster mass. The cluster is created instantaneously, i.e. every star has the same age. This is designed to model young, compact clusters. In Section \ref{discussion} we discuss how relevant this approximation is to observed structured giant molecular clouds.

\subsection{Stellar Evolution Model}
\label{model:stellarevolution}

To calculate the quantities emitted by the stellar sources (e.g. wind momentum deposition rate, ionising photon emission rate, ionising photon energy), we use the stellar evolution model described in \cite{Geen2018}, with a wind model based on \cite{Gatto2017}. We use the rotating tracks given in the Geneva tables \citep{Ekstrom2012}, with spectra calculated in \textsc{Starburst99} \citep{Leitherer2014} using these tracks. The surface temperature $T_{*}$ for each star is taken from the Geneva tables and used to calculate the ionised gas temperature in Section \ref{model:ionisedgasproperties}.

We interpolate linearly between model tracks of different stellar masses. During the interpolation, each track is normalised to a time between 0 and 1, where 0 is the start of the Zero Age Main Sequence (ZAMS) and 1 is the lifetime of the star. In this way, we optimise the ability for interpolated tracks to retain features in the neighbouring model tracks that scale with the lifetime of the stars. In Sections \ref{results:single-stars}, \ref{results:small-clusters} and \ref{results:lowmetal}, where not otherwise specified, we assume a stellar age of 1 Myr.

\subsection{Ionised Gas Properties}
\label{model:ionisedgasproperties}

\begin{figure}
	\centerline{\includegraphics[width=0.95\columnwidth]{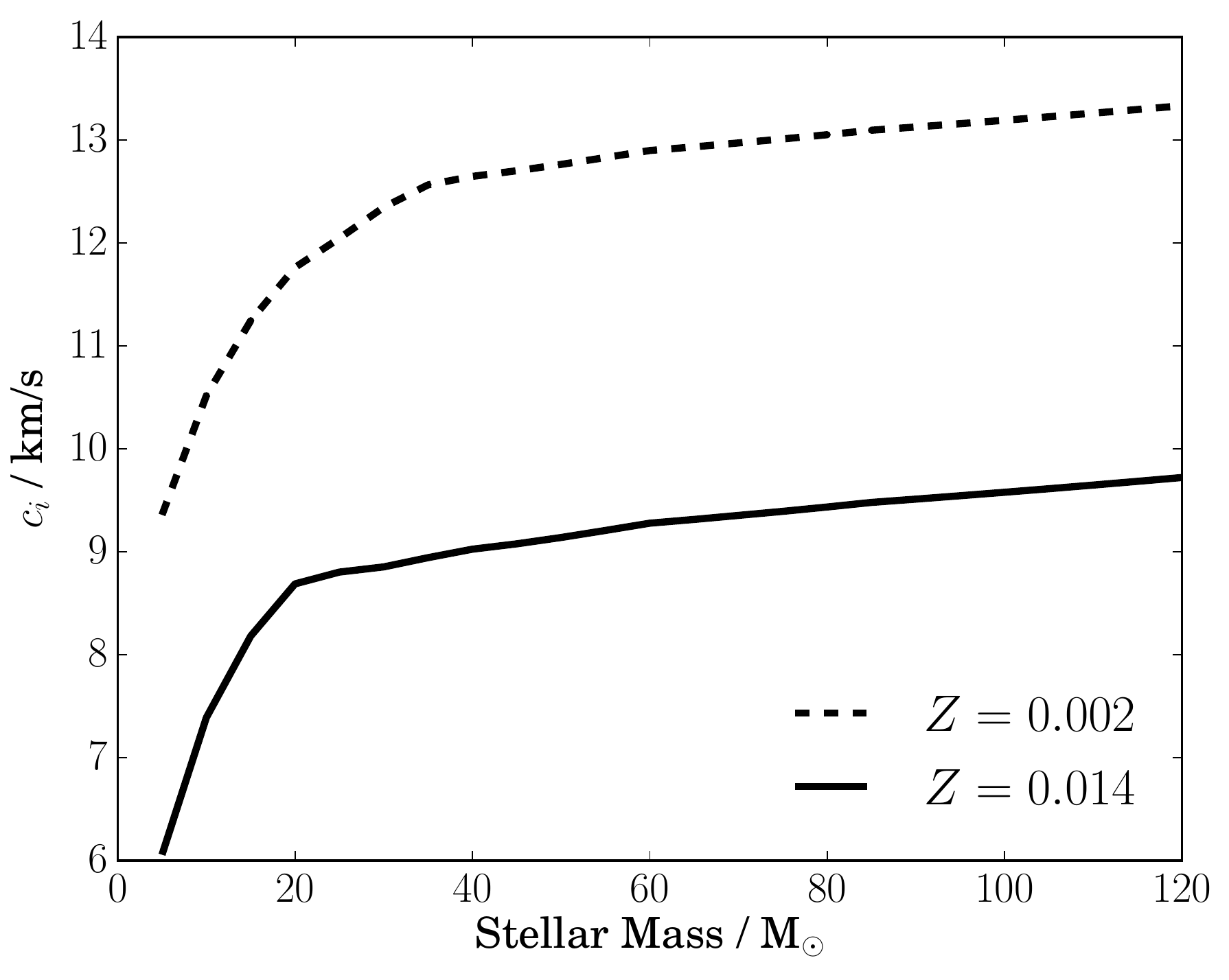}}
	\caption{Ionised gas sound speed $c_i$ in \hii regions around each star of age 1 Myr with metallicities $Z=0.002$ and $Z=0.014$. See Section \ref{model:ionisedgasproperties}.}
	\label{fig:ionisedsoundspeed}
\end{figure}

In order to obtain reliable values for $c_i$, we use the spectral synthesis code Cloudy \citep{Ferland2017}. We generate a table of ionised gas sound speeds $c_i$ for a set of time-dependent stellar surface temperatures $T_{*}$, as well as gas density $n_i$, metallicity $Z$ and ionisation parameter ${\cal U}$. $T_*$ is taken from the Geneva tables as described in Section \ref{model:stellarevolution}. Where multiple stars are present (see Section \ref{results:small-clusters}), the temperature of the hottest star is used. To remove degeneracies in the solution, we pick a fiducial value of $n_i=50$ \atcc and ${\cal U} =-2$. Although $n_i$ varies with time, it has a relatively weak effect on $c_i$. \rev{These values are plotted in Figure \ref{fig:ionisedsoundspeed}.}

To calculate $\alpha_B$, we use the temperature-dependent model given in, e.g., Appendix E2 of \cite{Rosdahl2013}. The value in all cases is close to $2 \times 10^{-13}$ cm$^3$ / s. We assume $\gamma = 4/3$.

\section{Results}
\label{results}

In this Section we present solutions to the models described in the previous Sections. We begin by showing how the radius and density of an \hii region in a specific environment evolves over time. We then calculate $C_w$, $C_{rp}$, $C_{B}$ and $r_{launch}$ in a set of physically-motivated conditions designed to explore where winds, radiation pressure, ionisation front breakout and gravity should become important. We solve these equations both for individual stars and for clusters sampled from an \IMF. Model and parameter choices are given in Section \ref{model}. We focus for the majority of this Section on solar metallicity results. In Section \ref{results:lowmetal} we present results for low metallicity stars and clouds.

\subsection{Evolution of an Example \hii Region}
\label{results:evolution-of-an-example-hii-region}

Here we introduce an example \hii region evolving under the influence of winds and UV photoionisation. We pick a 120~\Msolar star in a $10^4$~\Msolar cloud bounded by a surface density of 100~\Msolarpc. This is equivalent to the Orion A and B clouds, as listed in \cite{Lada2010}. However, this choice requires that these clouds have a single peaked density profile rather than multiple density peaks, which is unrealistic. We choose this set of conditions largely to demonstrate a stronger response to winds and radiation pressure.

\begin{figure}
	\centerline{\includegraphics[width=0.95\columnwidth]{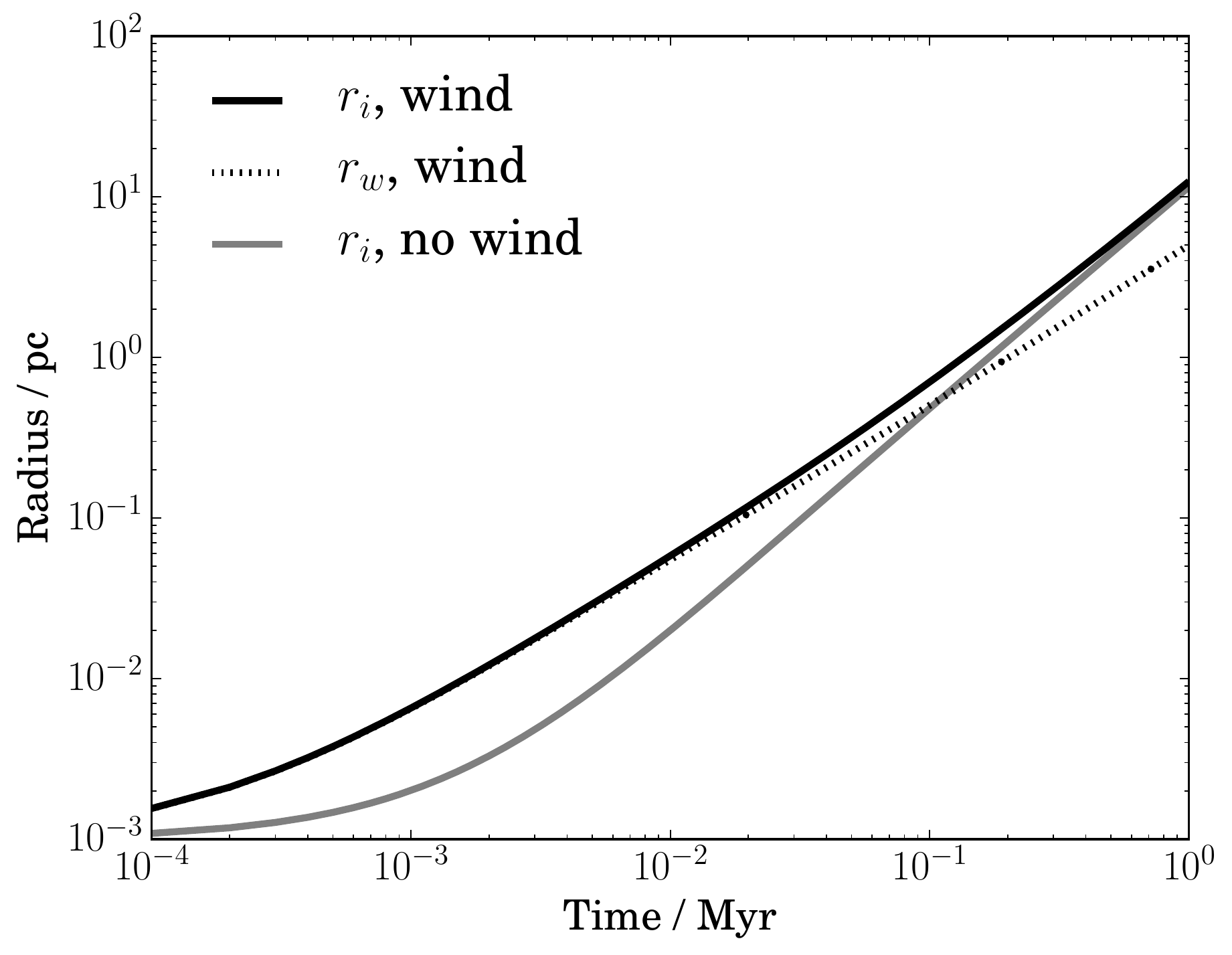}}
	\centerline{\includegraphics[width=0.95\columnwidth]{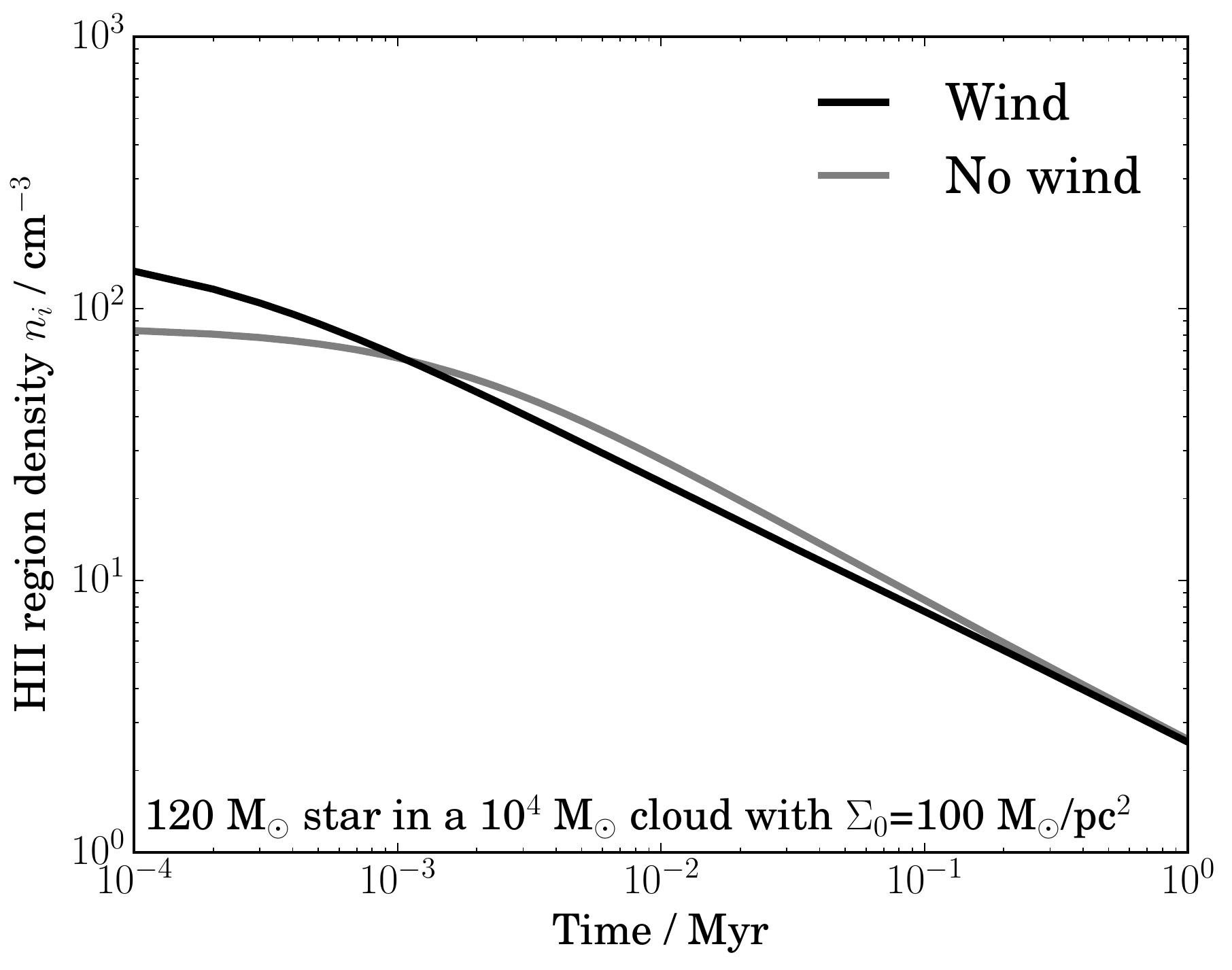}}
	\centerline{\includegraphics[width=0.95\columnwidth]{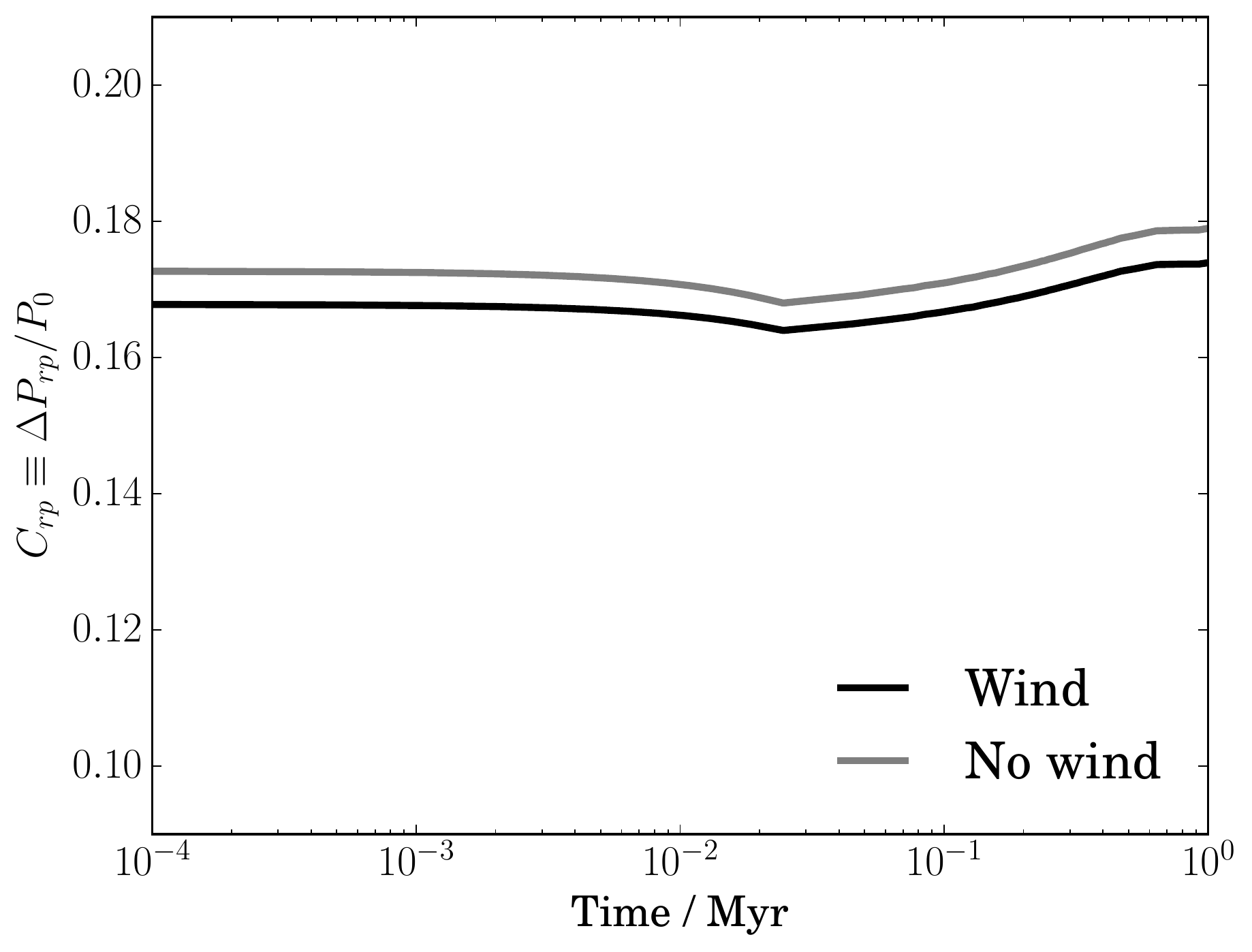}}
	\caption{Plots showing the evolution of a single 120 \Msolar star in a $10^4$ \Msolar cloud bounded by a surface density of 100 \Msolarpc. The top plot shows the evolution of the wind and ionisation front radii $r_w$ and $r_i$ as given in Equation \ref{wind:dynamics}. The middle plot shows the density of the \hii region $n_i$ in the same solution as given in Equation \ref{wind:density} in solutions both including and ignoring stellar winds. The bottom plot shows the relative effect of radiation pressure and thermal pressure on the expansion of the \hii region $C_{rp}$ in Equations \ref{radiation_pressure:crpfull} and \ref{radiation_pressure:crpsimple}. We start each solution at $r=0.001$ pc since the photoionisation-only equations have no defined solution if they start at $r=0$.}
	\label{fig:singlestar_evolution}
\end{figure}

Figure \ref{fig:singlestar_evolution} shows the results of integrating the solutions to the equations in Sections \ref{winds_in_uv} and \ref{radiation_pressure}, \rev{using the same parameters as introduced in these Sections (i.e. an efficiently cooling, momentum-driven wind, and radiation pressure terms where $f_{trap}=2$, $\phi_d=0.73$ and $\psi=1$)}. We start from a radius of 0.001 pc and not 0 pc, since a) in reality protostellar outflows set the initial radius of the \hii region \cite[see][]{Kuiper2018}, and b) the ionisation front cannot expand unless $r_i$ is larger than the bistable point $r_{launch}$. 

The initial expansion of the \hii region is faster in the case where winds are included, though at large radii there is little difference between the solutions. The wind bubble occupies most of the volume of the \hii region at small radii, while at 20 pc the wind bubble radius $r_w$ is only 20\% of the \hii region radius $r_i$.

The winds initially compress the \hii region by a small factor, before causing its density to drop compared to a solution without winds. As with the radius, at later times the results with and without winds converge.

\rev{The contribution to the \hii region's pressure from radiation pressure is between 16 and 18\% in this model. Photoionisation is thus more important, but radiation pressure is significant enough to suggest that more detailed calculations with radiation pressure and winds included self-consistently should be considered for a complete model of the \hii region. The small variations in $C_{rp}$ in the bottom panel of Figure \ref{fig:singlestar_evolution} are due to changes in the surface temperature of the star, which in turn affects the equilibrium temperature of the photoionised gas.}

\rev{When we include winds in the calculation, $C_{rp}$ drops, i.e. winds reduce the impact of radiation pressure in this instance. This is due to the behaviour of Equation \ref{radiation_pressure:crpfull}, where $C_{rp} \propto r_i^{-2} n_i^{-1}$. Radius $r_i$ is larger at all times in the solution with winds (top panel in Figure \ref{fig:singlestar_evolution}), and density $n_i$ is initially larger (middle panel). This is because the wind bubble acts like a piston on the inside of the \hii region, pushing and compressing it. Thus the overall effect of winds is to reduce $C_{rp}$.}

Having illustrated how one system behaves, we now explore how the effects of winds and radiation pressure change over a larger range of stellar and cloud parameters.

\begin{figure*}
	\centerline{\includegraphics[width=0.95\columnwidth]{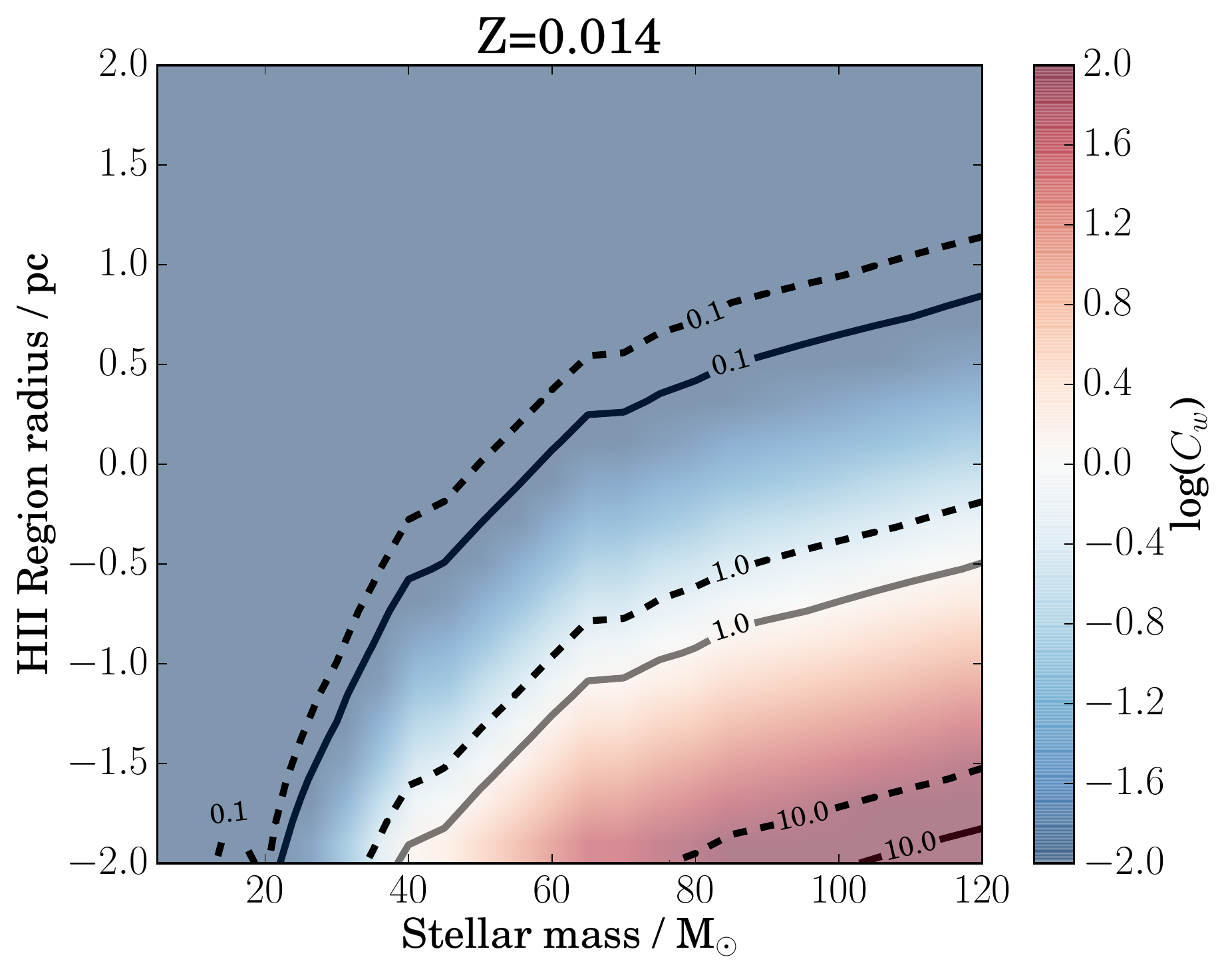}
	            \includegraphics[width=0.95\columnwidth]{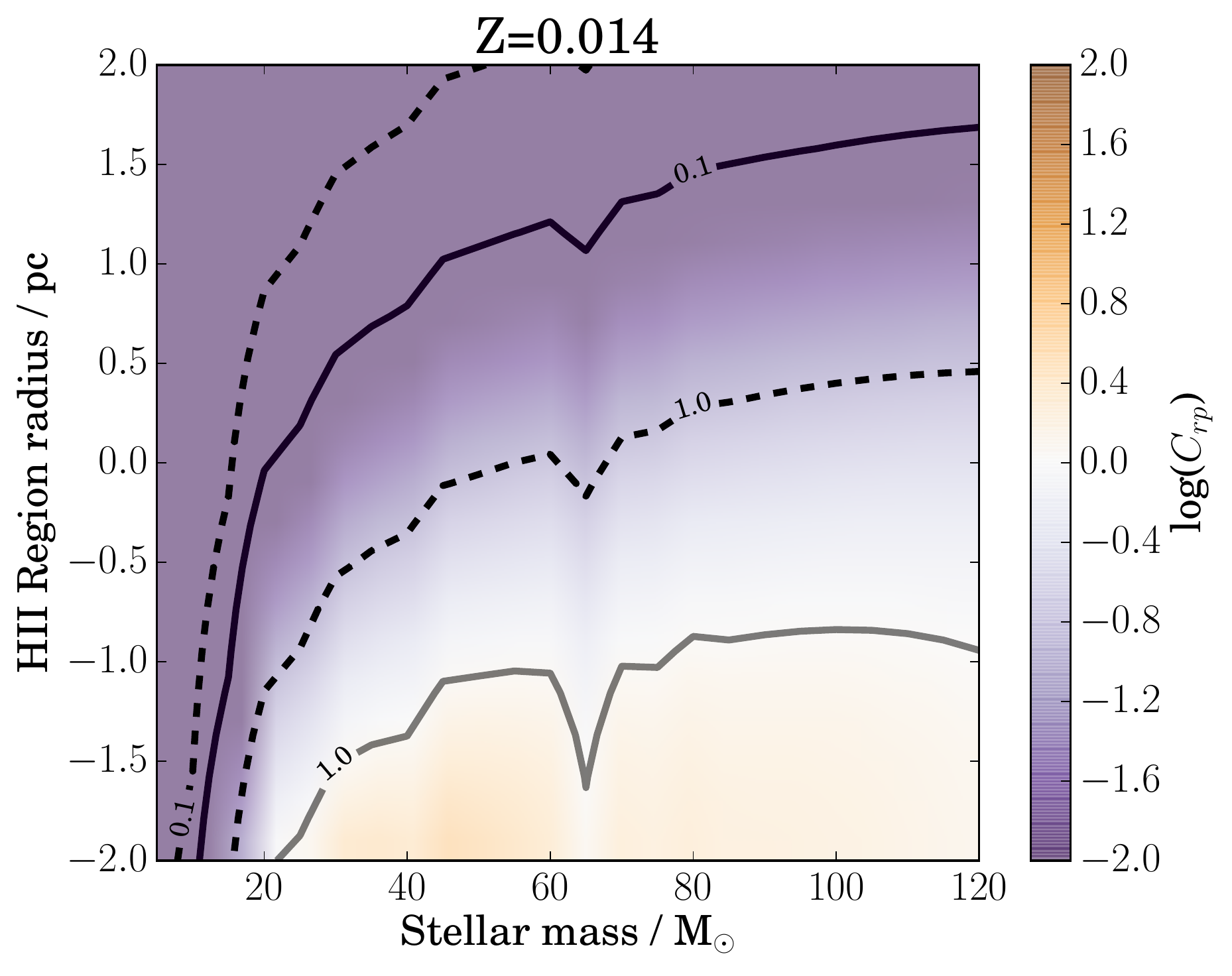}}
	\caption{Surface plots of condition $C_w$ for winds and $C_{rp}$ for radiation pressure for \hii regions of a given radius around single stars. A value of $> 1$ means that winds or radiation pressure have a large influence on the dynamics of the \hii region that cannot be ignored. Solid lines show results for a single star of a given mass and age 1 Myr on the x-axis, dashed lines for a pair of stars each with the given mass. The slight dip in $C_{rp}$ at 65 \Msolar is due to a change in the metal line absorption at this mass. The sound speed in the photoionised gas and other parameters used in these calculations discussed in Section \ref{model}.}
	\label{fig:surface_radius}
\end{figure*}

% Surface plots showing Cw and Crp
\begin{figure*}
	\centerline{
		\includegraphics[width=\columnwidth]{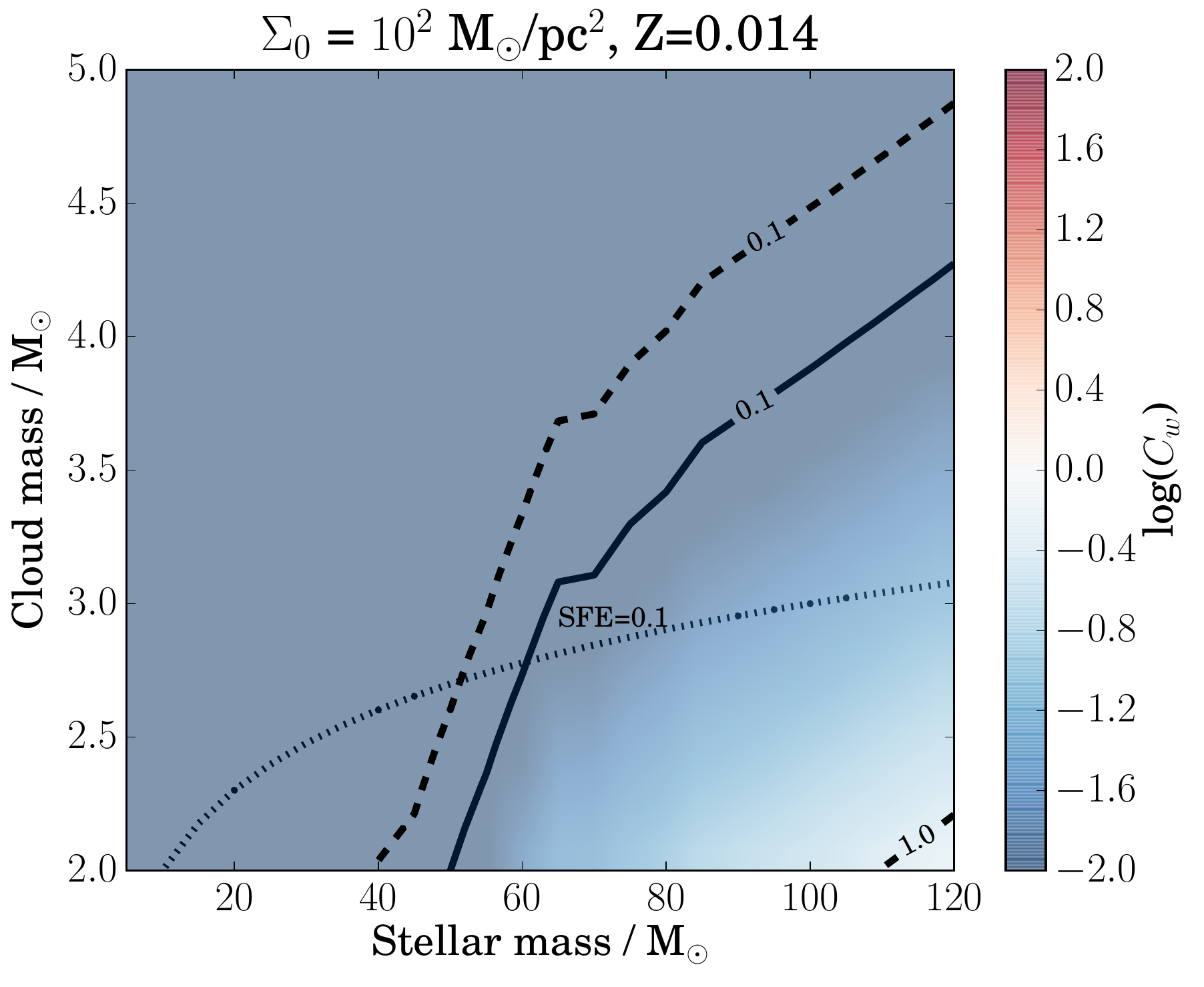} 
		\includegraphics[width=\columnwidth]{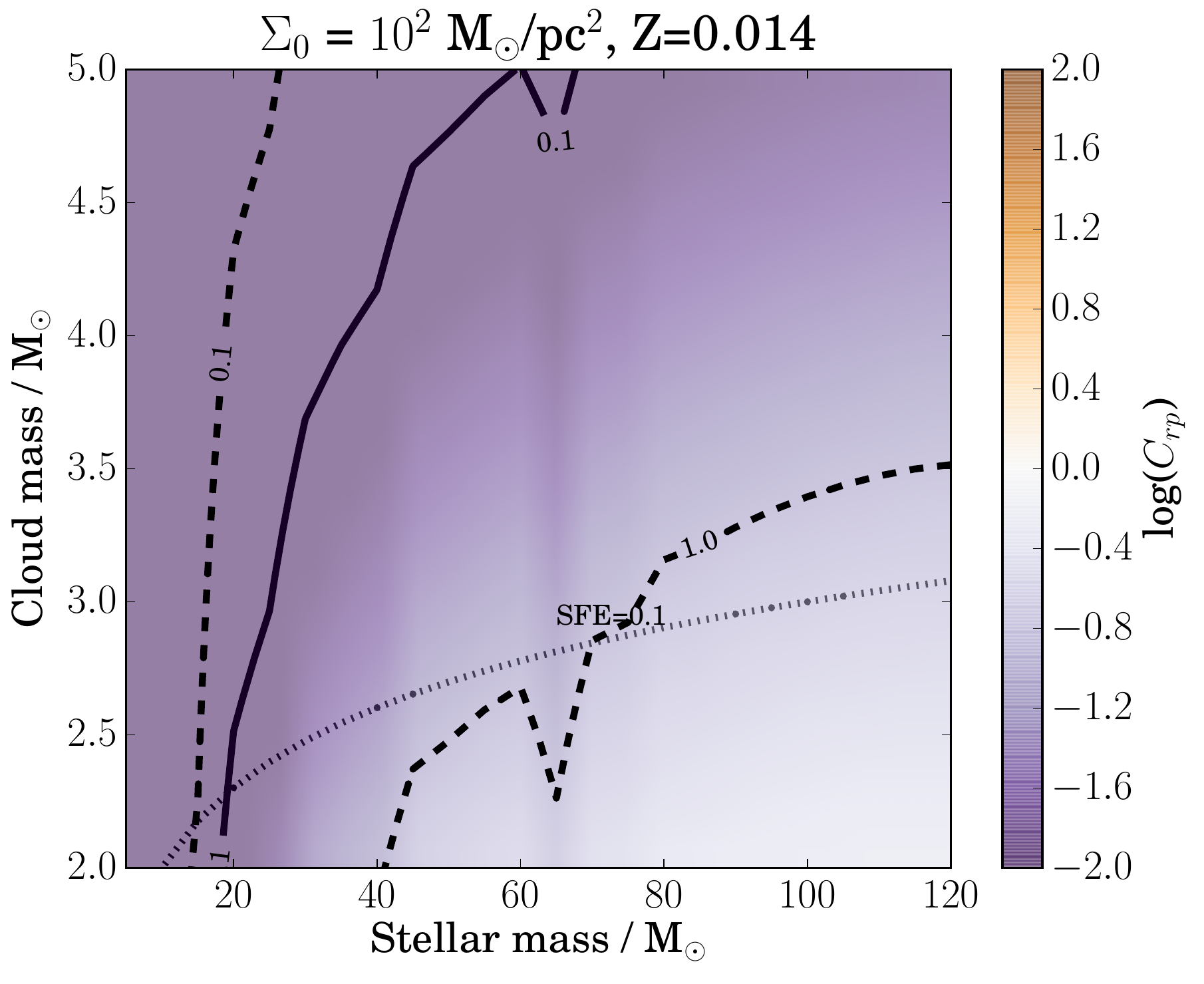}
	}
	\centerline{
		\includegraphics[width=\columnwidth]{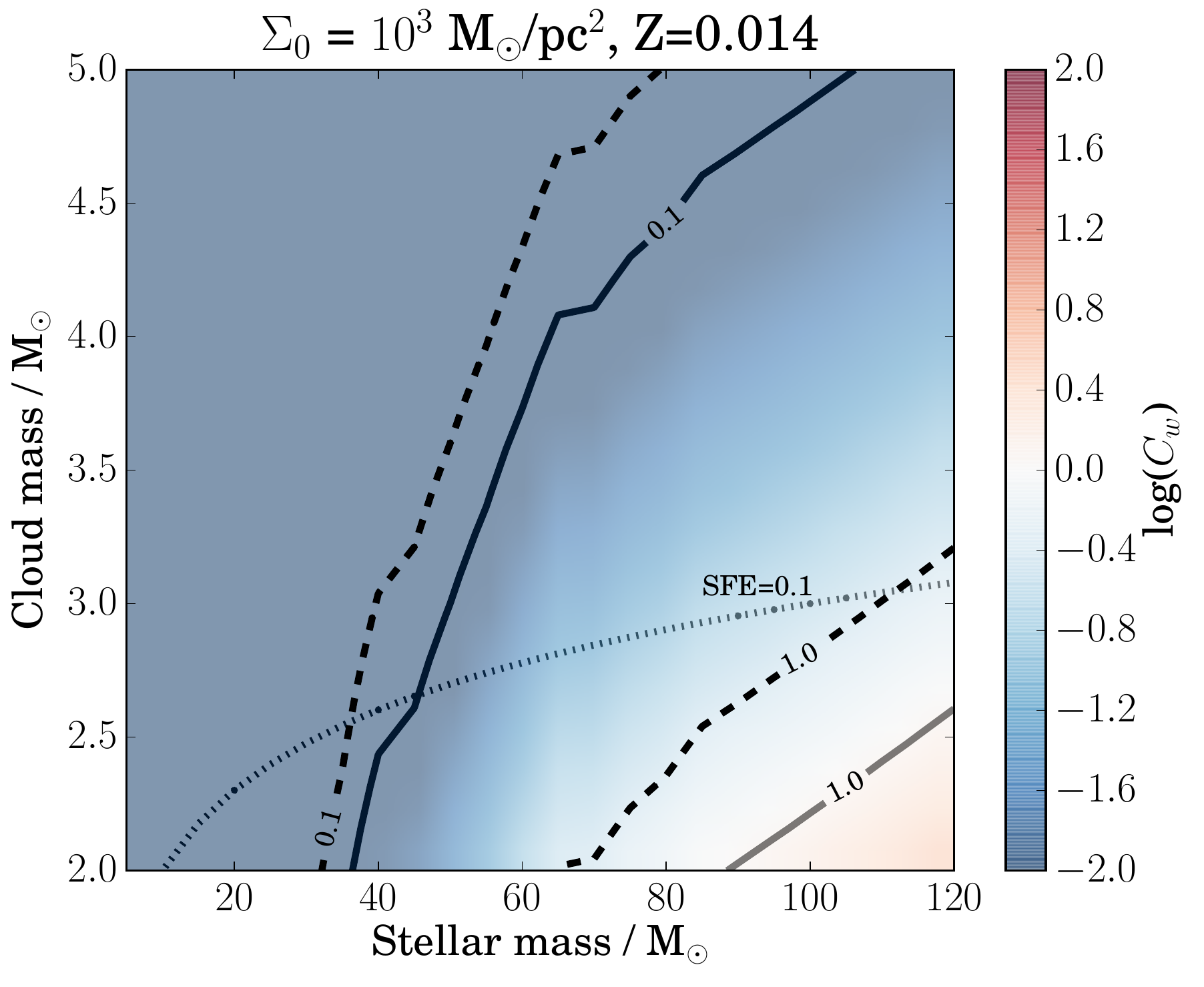} 
		\includegraphics[width=\columnwidth]{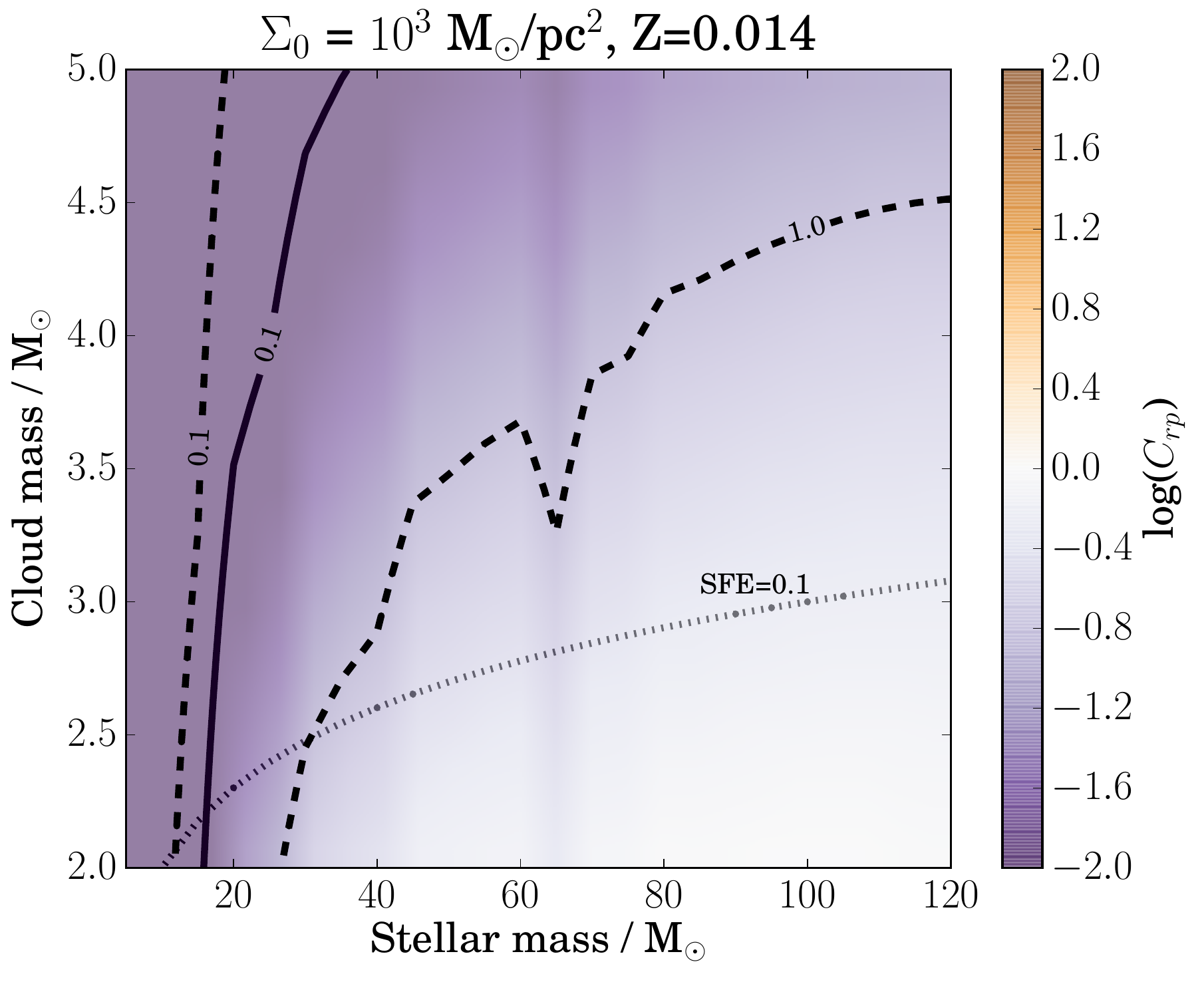}
	}
	\centerline{
		\includegraphics[width=\columnwidth]{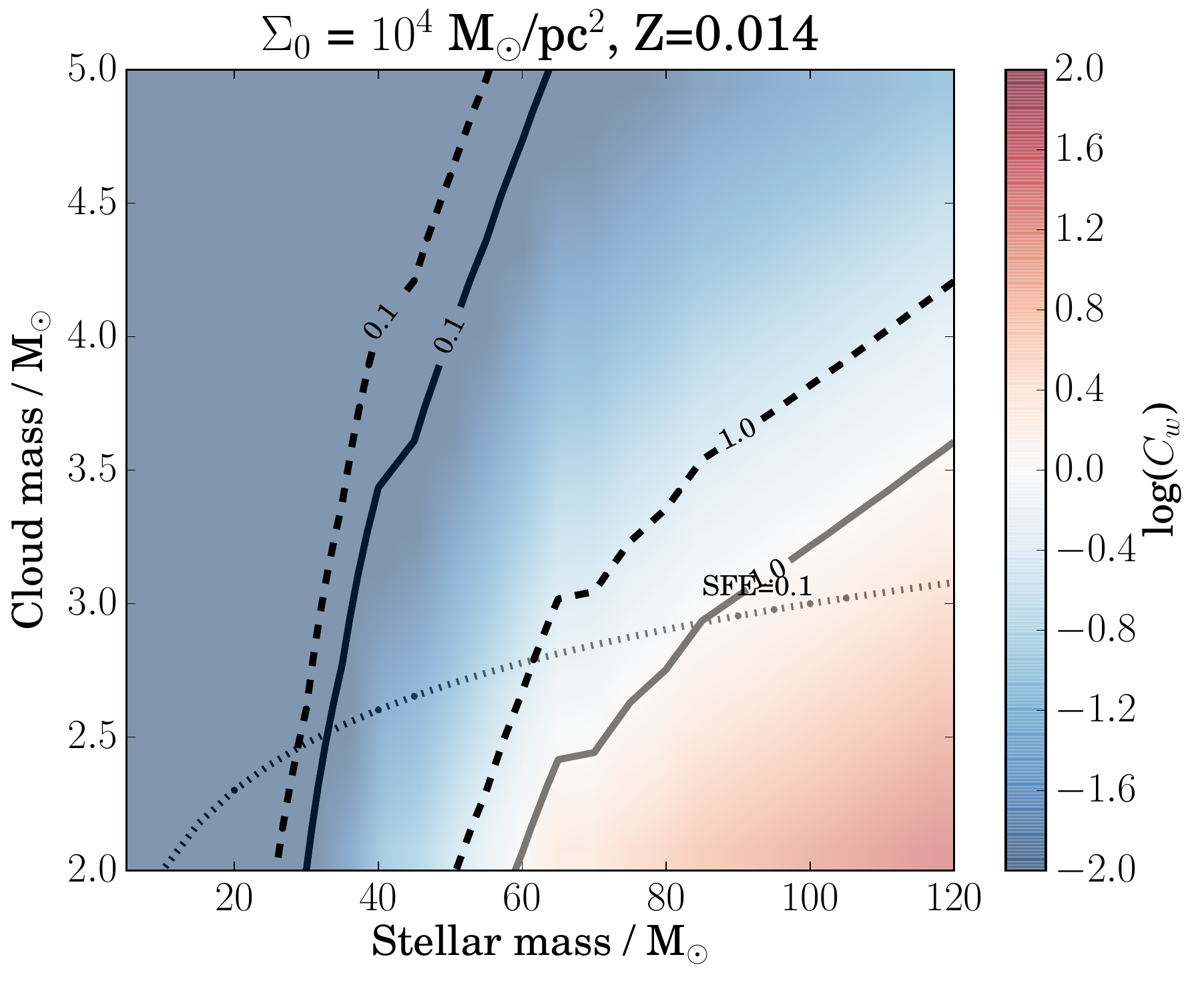} 
		\includegraphics[width=\columnwidth]{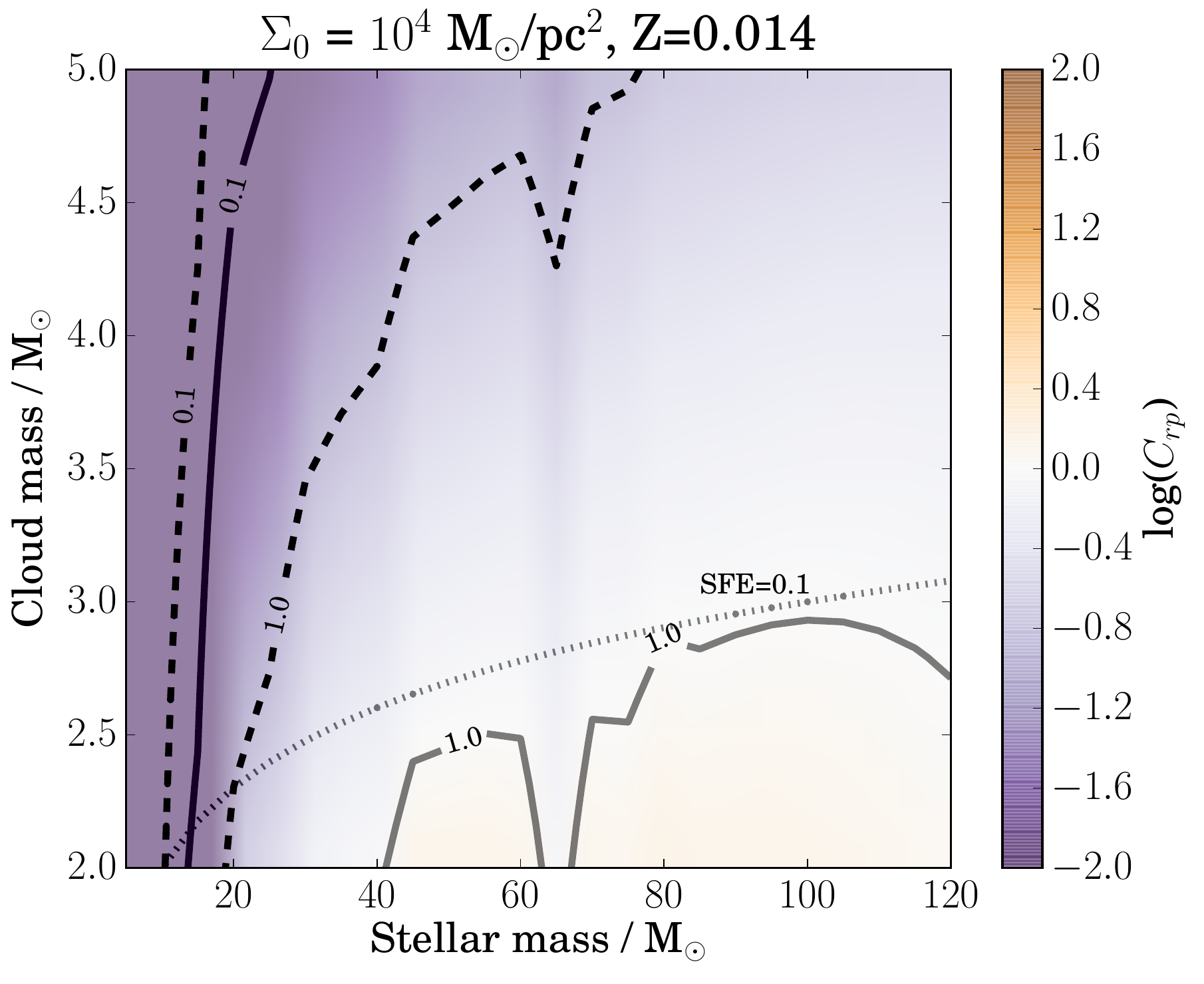}
	}
	\caption{Surface plots of conditions $C_w$ and $C_{rp}$ at the cloud edge for single stars of age 1 Myr. A value $> 1$ means that winds or radiation pressure exert more influence over the \hii region than photoionisation. Solid lines show results for a single star of a given mass on the x-axis, dashed lines for a pair of stars each with the given mass. Values are given at radius $r_0$ where the bounding surface density is $\Sigma_0$ and the initial enclosed mass is $M_0$ (see Section \ref{cloud}). \rev{A dotted line shows cloud masses $10\times$ the star's mass to give an indication of where cloud masses at this ratio lie in the parameter space.}}
	\label{fig:surfaceplots}
\end{figure*}

\begin{figure*}
	\centerline{\includegraphics[width=0.95\columnwidth]{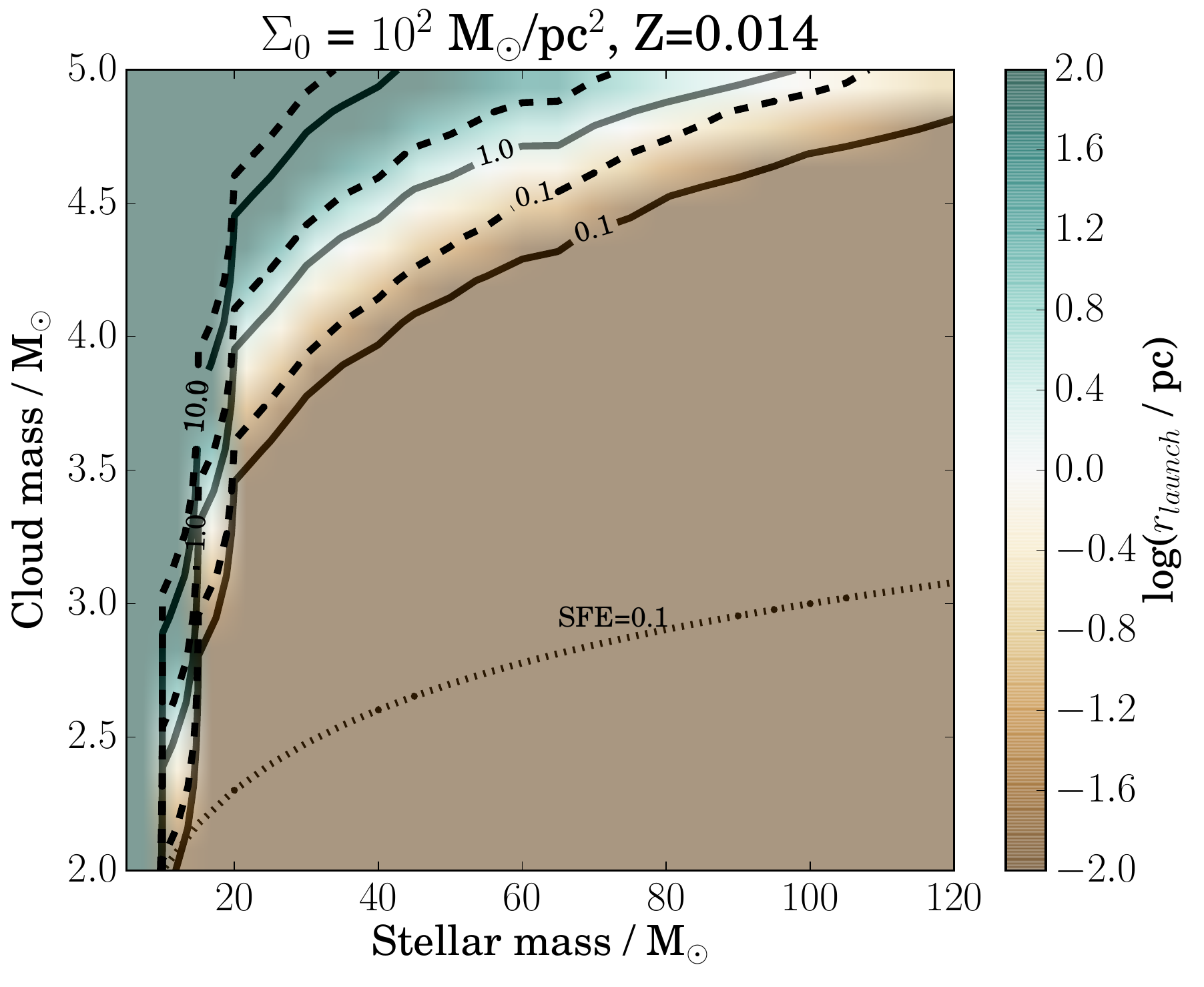}
		\includegraphics[width=0.95\columnwidth]{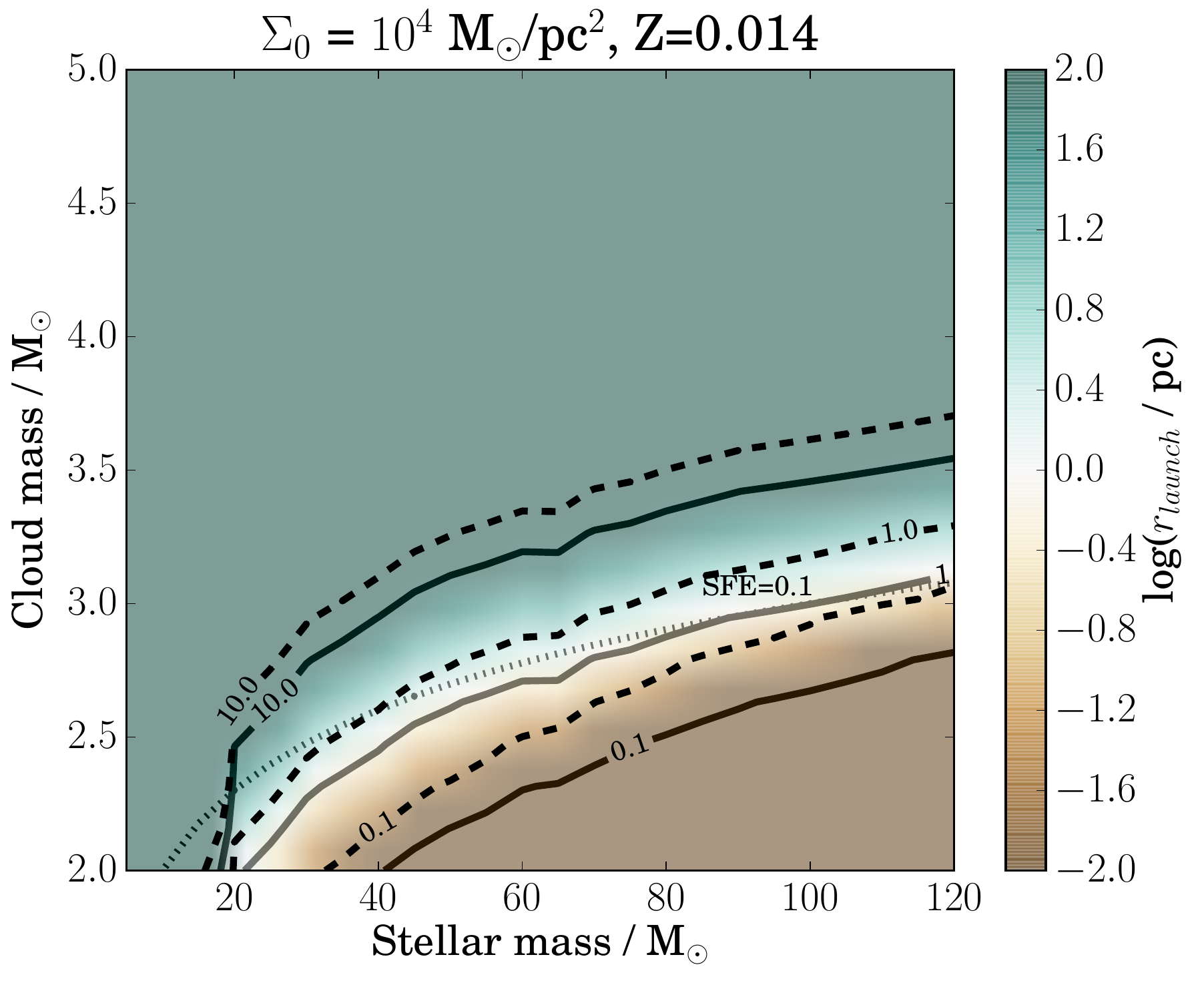}}
	\caption{Surface plots of launch radius in pc for single stars of age 1 Myr at various cloud masses, enclosed by surface densities of 100 and 10000 \Msolarpc, as in Figure \ref{fig:surfaceplots}. Note that for steep density profiles, a \textit{small} $r_{launch}$ is required to allow the ionisation front to expand. Solid lines show results for a single star of a given mass on the x-axis, dashed lines for a pair of stars each with the given mass. A dotted line shows cloud masses $10\times$ the star's mass to give an indication of where cloud masses at this ratio lie in the parameter space.}
	\label{fig:rstall}
\end{figure*}

\begin{figure*}
	\centerline{\includegraphics[width=0.95\columnwidth]{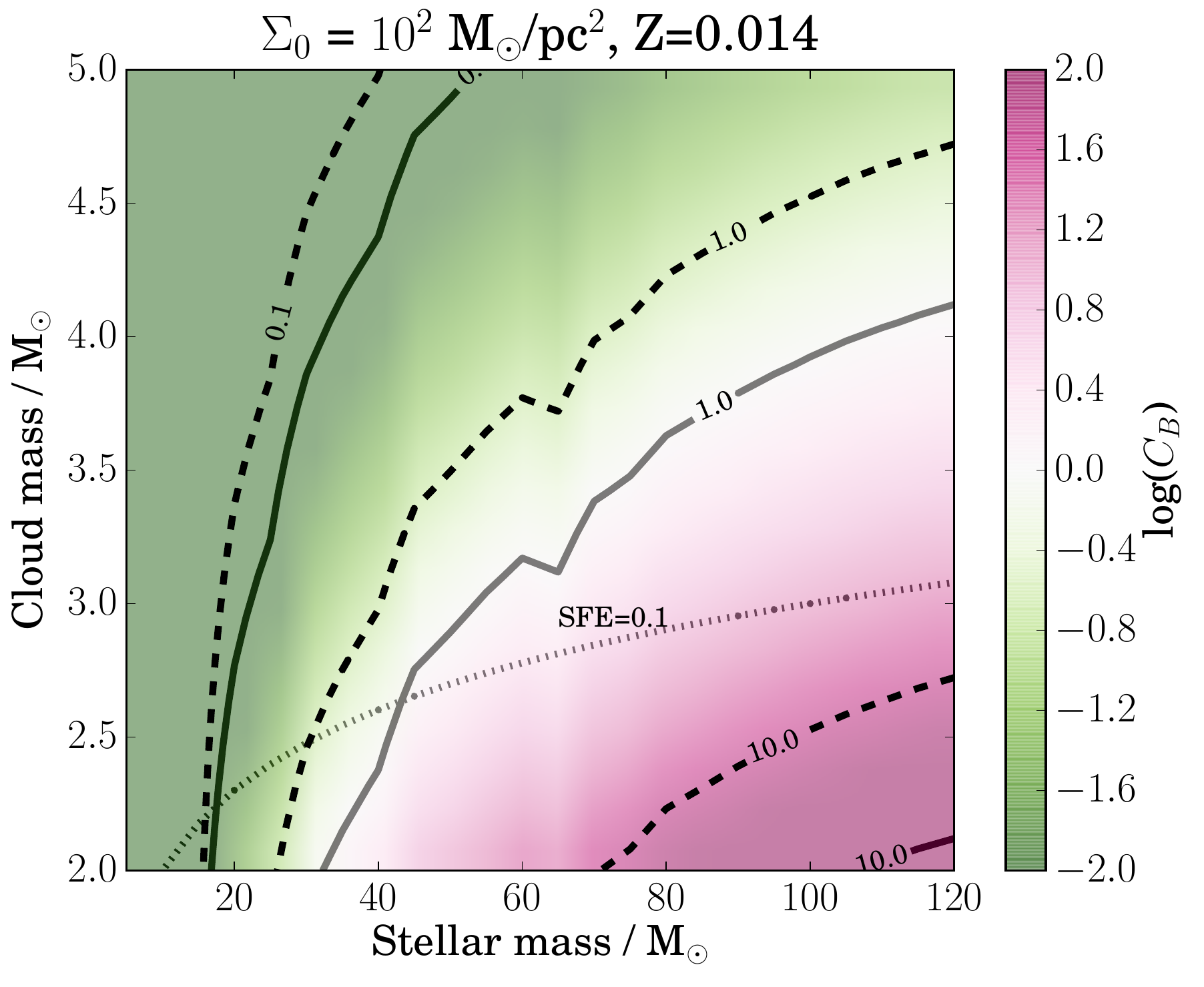}
		\includegraphics[width=0.95\columnwidth]{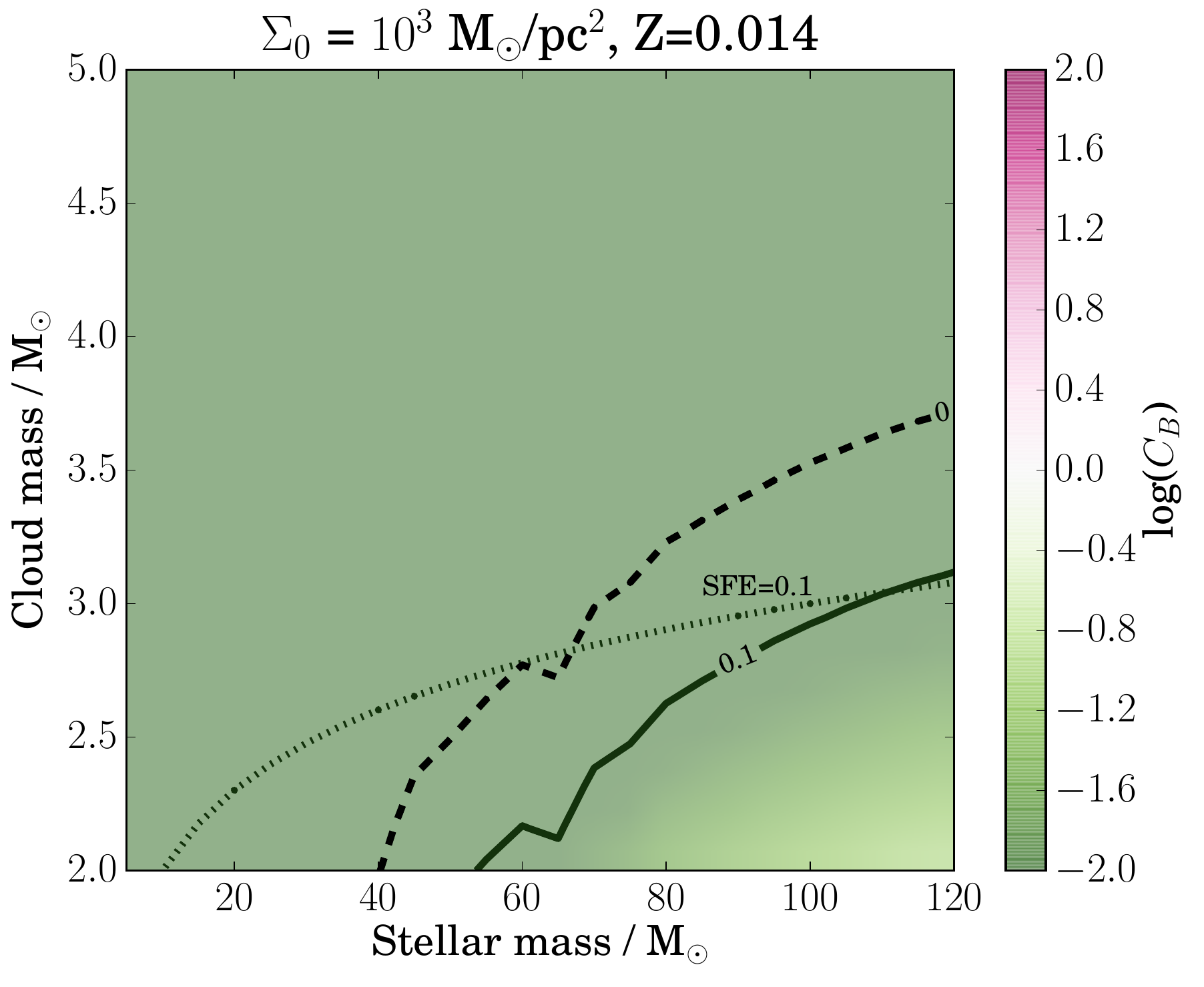}}
	\caption{Surface plots of the breakout condition $C_{B}$ for single stars of age 1 Myr at various cloud masses, enclosed by surface densities of 100 and 1000 \Msolarpc. $C_{B}>1$ implies that the ionisation front should break out of its shell as calculated by \protect\cite{Alvarez2006}. Solid lines show results for a single star of a given mass on the x-axis, dashed lines for a pair of stars each with the given mass. A dotted line shows cloud masses $10\times$ the star's mass.}
	\label{fig:breakout}
\end{figure*}

\subsection{Single Stars and Pairs}
\label{results:single-stars}

In Figure \ref{fig:surface_radius}, we plot $C_w$ and $C_{rp}$ for \hii regions of a given radius around single stars, with contours for pairs of equal-mass stars representing non-interacting binaries overlaid. These coefficients determine whether winds or radiation pressure significantly affect the dynamics of the \hii region, respectively. Based on these results, winds and radiation pressure around single stars and binaries are most important at small radii, i.e. in the Ultracompact \hii region stage. The results for winds are consistent with the calculations of \cite{Capriotti2001}, who study a uniform external medium (i.e. $w=0$, rather than $w=2$ in this work). \rev{The use of equal-mass binaries gives winds a small boost, but radiation pressure a significantly larger boost. This is significant because, as \cite{Sana2012} argue, most massive stars are in binary pairs. As in Section \ref{results:evolution-of-an-example-hii-region}, even when $C_{rp}$ is less than 1, winds and radiation pressure combined have non-linear effects on the \hii region and should be treated carefully.}

Since we assume a quasi-hydrostatic internal structure for the \hii region, $C_w$ and $C_{rp}$ are time-independent. However, the properties of the star(s) will affect the values obtained. We assume a young massive star of age 1 Myr, since we are principally interested in the early expansion of the \hii region into the cloud. For very old clusters, the role of radiation (versus later kinetic processes such as supernovae) is reduced as the most massive stars die out. See \cite{Rahner2017} for a discussion of this phase and the relative importance of each feedback process.

In Figure \ref{fig:surfaceplots} we plot $C_w$ and $C_{rp}$ against cloud mass $M_0$ and bounding surface density $\Sigma_0$, where $r_i = r_0$. Note for a given gas density field $n(r_i) \propto r_i^{-2}$, as $r_i$ expands and encloses more mass, $M(<r_i)$ will increase and $\Sigma(r_i)$ will decrease. Winds are more important in very dense environments, either closer to the star or in globally denser clouds. Radiation pressure contributes at the 10\% level over a larger set of conditions, but rarely becomes more significant than photoionisation \rev{except in dense environments or around equal-mass pairs of stars (again, this does not account for the role of interacting binaries in stellar evolution)}.

Figure \ref{fig:rstall} shows the effect of gravity and addresses the question of whether the ionisation front should be trapped by self-gravity of the cloud and/or accretion flows. Note that as in the previous Section, a \textit{small} $r_{launch}$ is required for the ionisation front to escape in steeply peaked molecular clouds where $w > 3/2$. For stars above 20 \Msolar in surface densities of $\Sigma_0=100$ \Msolarpc, $r_{launch}$ is smaller than 0.1~pc, and so provided the ionisation front can reach this point (via protostellar outflows, non-spherical geometries, or some other effect) it can readily accelerate outside the cloud. For very dense clouds, the ionisation front needs to reach at least 0.1~pc for it not to be crushed by gravity. In this case, more details about the small-scale stellar and cloud physics that sets the initial \hii region radius are needed to form a more concrete model for this phase.

Figure \ref{fig:breakout} shows the condition for ionising photons to break out of the D-type \hii region and flood into the \ISM, i.e. $C_B>1$. This is possible for stars or binary pairs totalling above 70 \Msolar in clouds bounded by $\Sigma_0=100$ \Msolarpc, assuming 10\% of the cloud mass is in stars. Since $C_B\sim1$ in this instance, the breakout is likely to be borderline rather than a strong divergence from the D-type solution. Once the breakout has occurred, the cloud is rapidly and fully ionised, and the density field expands at a few times the speed of sound in the ionised gas \citep{Franco1990}. For denser clouds, breakout is not possible. Note again that we do not discuss shell fragmentation here \citep[see][]{Rahner2019}.

\subsection{Small Clusters}
\label{results:small-clusters}

As the \hii regions around individual stars merge, they form a larger \hii region that expands through the cloud as a whole. In this Section we discuss models where this occurs. We assume, in the case where winds exist, that the wind bubbles merge. If they do not, then the effect of wind is significantly reduced. This case is discussed in some detail by \cite{Silich2017}. It also assumes that winds cool efficiently, which we discuss in Section \ref{winds_in_uv}.

\begin{figure}
	\centerline{\includegraphics[width=0.95\columnwidth]{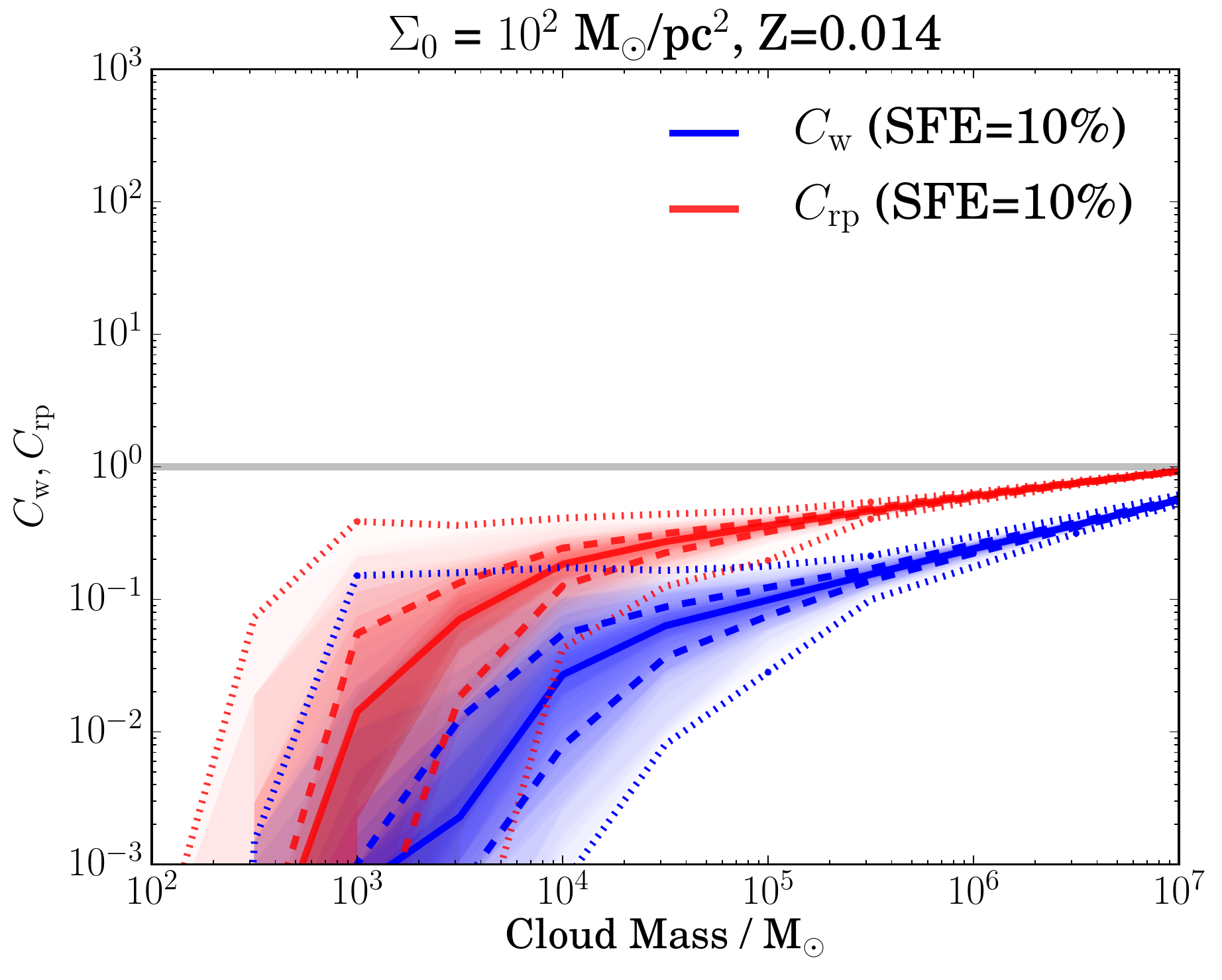}}
	\centerline{\includegraphics[width=0.95\columnwidth]{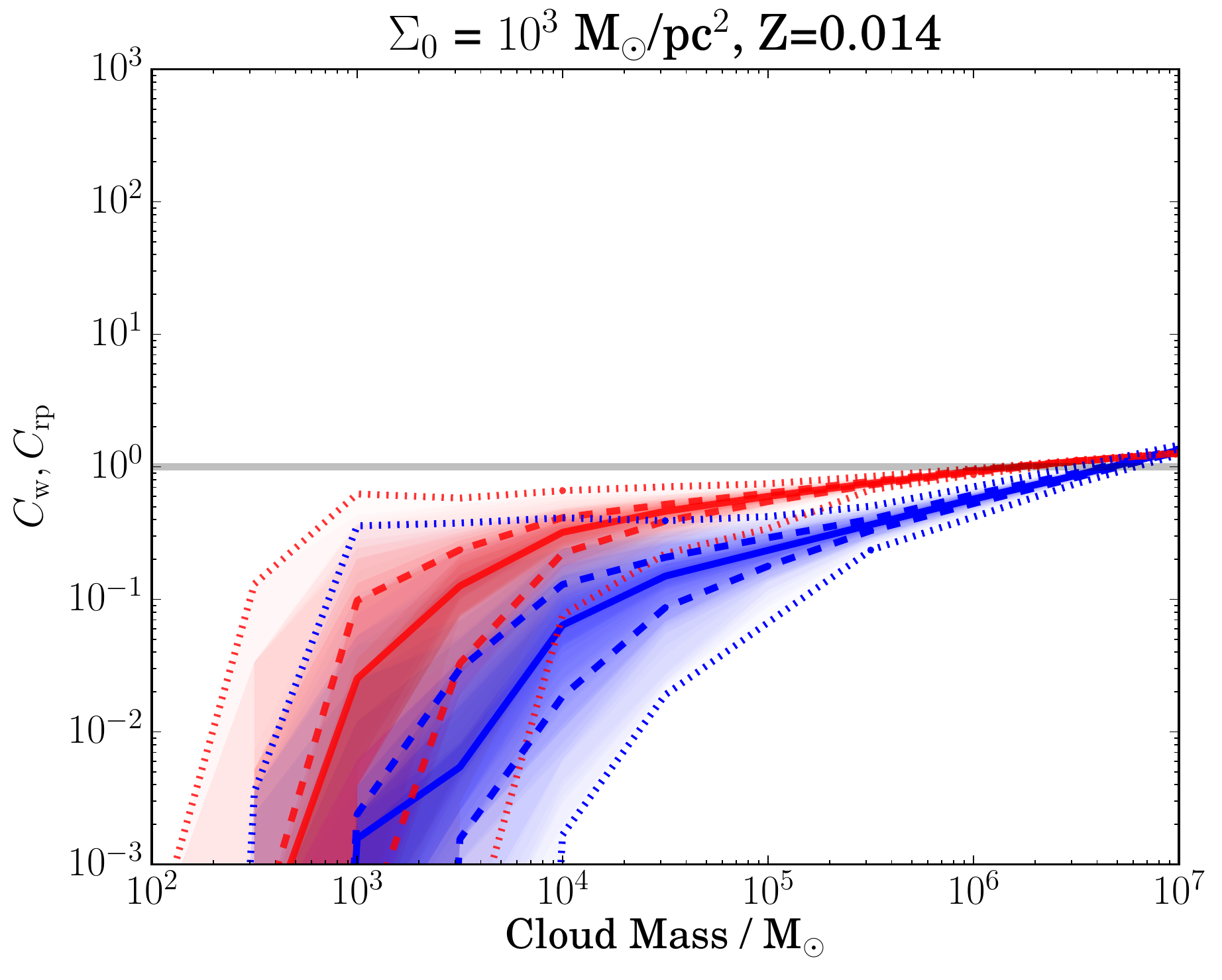}}
	\centerline{\includegraphics[width=0.95\columnwidth]{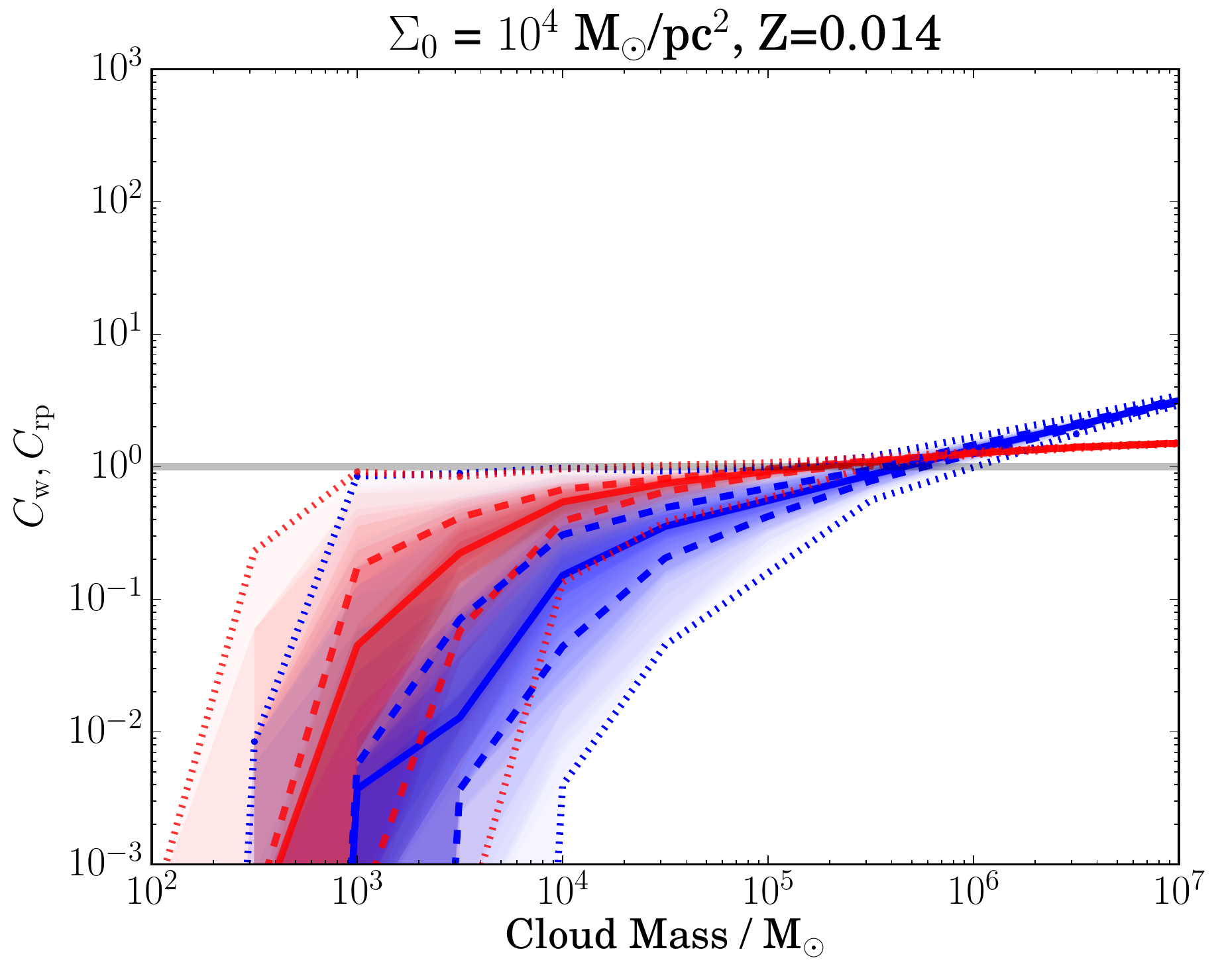}}
	\caption{Surface plots showing conditions $C_w$ and $C_{rp}$ for clouds and clusters at various masses and surface densities, where values $>$ 1 indicate that the effect is more significant than the pressure from UV photoionisation. The masses of stars in each cluster are randomly sampled from a \protect\cite{Chabrier2003} IMF up to a mass of 10\% of the cloud mass. $C_w$ is shown in red and $C_{rp}$ in blue. Dotted lines show the maximum and minimum values for each cluster mass, dashed lines show the interquartile range and solid lines the median. The shading shows every percentile, where denser colour indicates a value closer to the median. }
	\label{fig:sfeplots}
\end{figure}

In Figure \ref{fig:sfeplots} we invoke a cluster at $r=0$ by randomly sampling from an IMF up to 10\% of the cloud mass (see Section \ref{model:cloudprops}). The cluster is assumed to be contained entirely within the \hii region and wind bubble, i.e. the interaction between wind bubbles as in \cite{Silich2017} is ignored.

As in the single stellar case, winds only have an effect at the highest surface densities and cloud masses. \rev{Radiation pressure ($C_{rp}$) is more dynamically important than winds ($C_{w}$) in most cases, and can even reach the same importance as photoionisation in certain conditions. Note, however, that this is not a full calculation of the role of radiation pressure, but an estimate of the perturbation that it gives to the solution. Where $C_{rp}$ and $C_{w}$ are large, future work should target these conditions with more realistic models. Full multi-wavelength radiative hydrodynamic calculations including accurate dust models and cooling are needed to determine the precise role of radiation pressure and winds on \hii regions.}

Broadly, we expect photoionisation to be the most significant driver of \hii regions except in dense or high-mass clouds. In these clouds, more careful modelling is required, particularly since the SFE is likely larger than 10\% in these enivronments due to the relative inefficiency of feedback and short free-fall times \citep[e.g.][]{Vazquez-Semadeni2018,Rahner2019}.

\begin{figure}
	\centerline{\includegraphics[width=0.95\columnwidth]{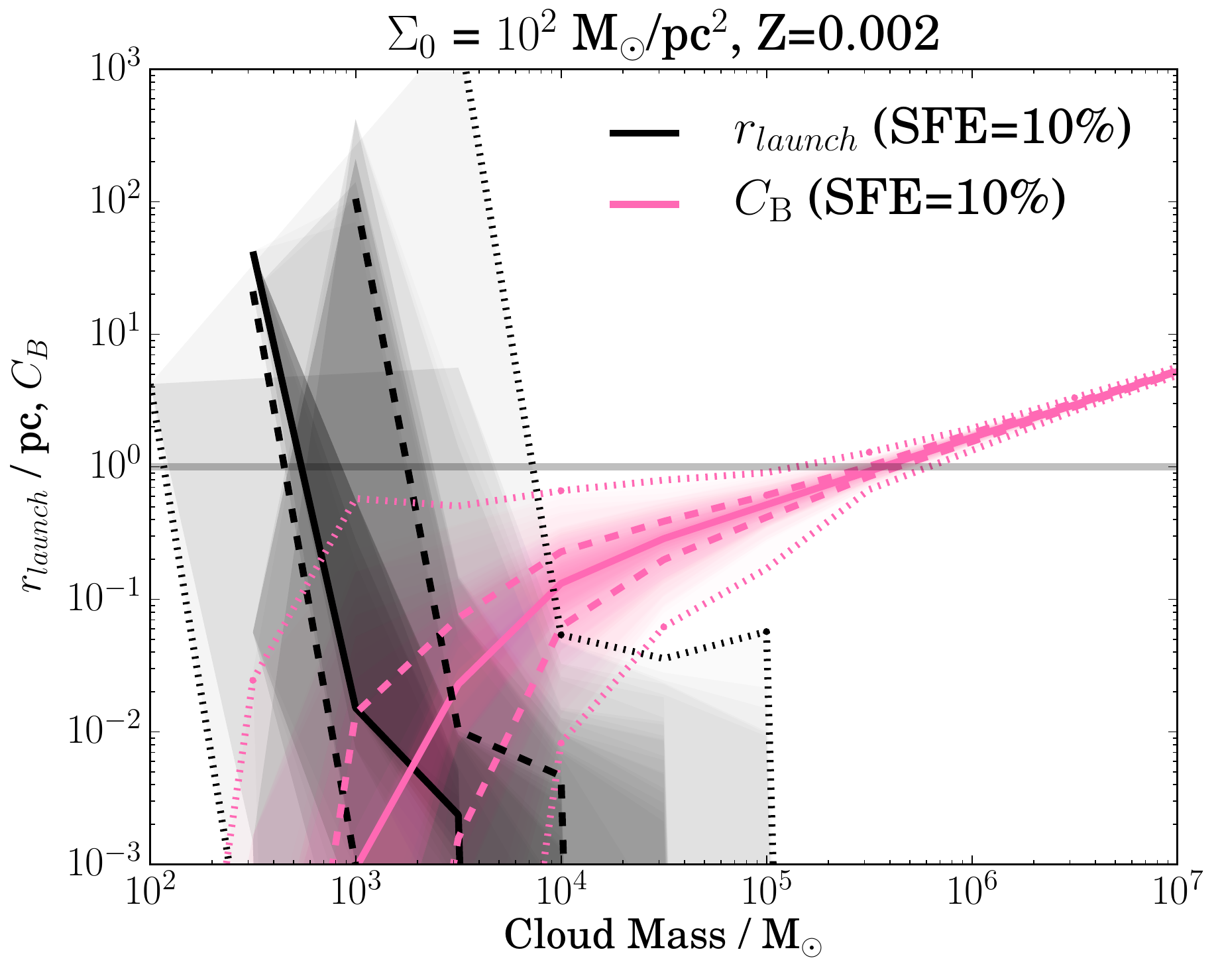}}
	\caption{Surface plots showing $C_B$ (the condition that ionising photons break out of D-type front) and $r_{launch}$ in pc for clouds and clusters at various masses and surface densities. Values are obtained as in Figure \ref{fig:sfeplots}. There is considerable scatter in the values for $r_{launch}$, but the main limiting factor to launch an \hii region is forming enough massive stars, which becomes ever more likely at higher cloud masses.}
	\label{fig:sfe_rstallcbplots}
\end{figure}

In Figure \ref{fig:sfe_rstallcbplots} we plot $r_{launch}$ and $C_B$ for the same clusters. For small clusters, insufficient photons are available to either drive the \hii region outwards or break out of the D-type solution. Between cloud masses of $10^3$ and $10^4$ \Msolar, the ability for an \hii region to form (i.e. $r_{launch}$ is small) depends on the sampling of the \IMF. Above this mass, the ionisation front is unlikely to stall.

For much larger clouds, photon breakout becomes likely. We caution again that very massive single power-law density profiles bounded by $\Sigma_0=100$ \Msolarpc are unlikely in conditions similar to the solar neighbourhood. Observed systems with this mass and bounding density are likely structures made up of multiple peaked cores distributed in space.

%This is a simple template for authors to write new MNRAS papers.
%See \texttt{mnras\_sample.tex} for a more complex example, and \texttt{mnras\_guide.tex}
%for a full user guide.
%
%All papers should start with an Introduction section, which sets the work
%in context, cites relevant earlier studies in the field by \citet{Others2013},
%and describes the problem the authors aim to solve \citep[e.g.][]{Author2012}.

\subsection{Low Metallicity Environments}
\label{results:lowmetal}

\begin{figure*}
	\centerline{\includegraphics[width=0.92\columnwidth]{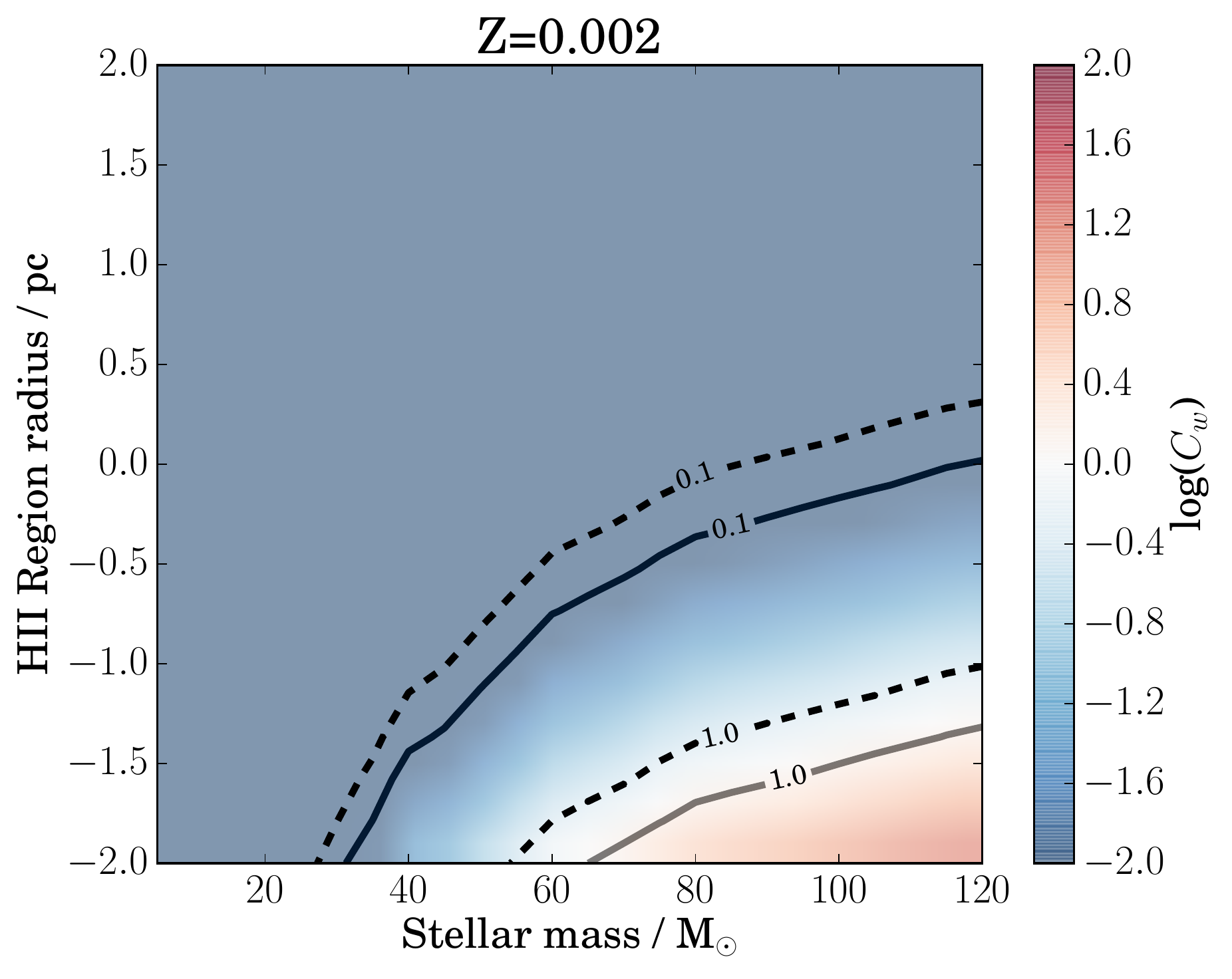}
		\includegraphics[width=0.92\columnwidth]{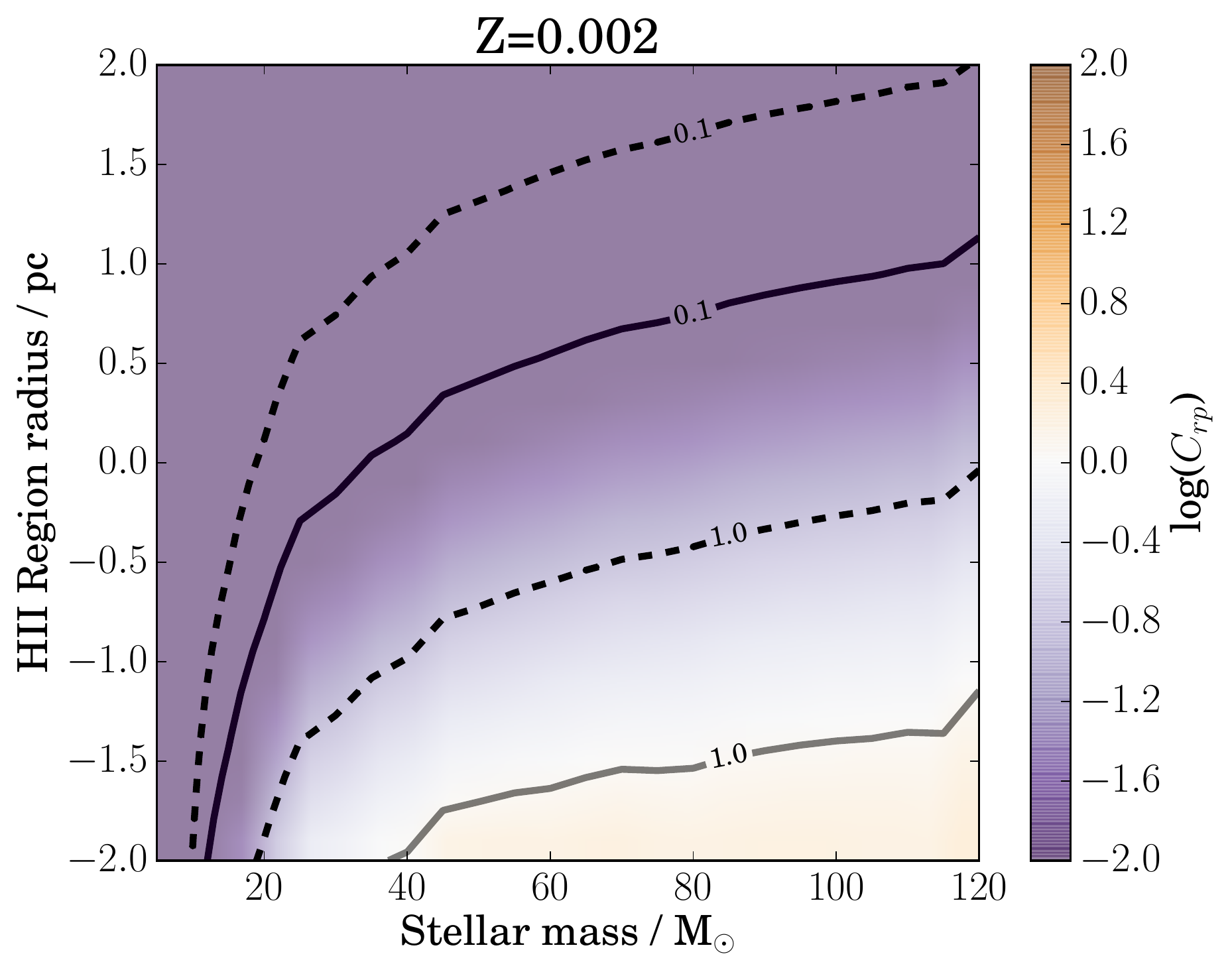}}
	\caption{Surface plots of condition $C_w$ for winds and $C_{rp}$ for radiation pressure at various radii at low metallicity ($Z=0.002$). See Figure \ref{fig:surface_radius} for solar metallicity results.}
	\label{fig:surface_radius_lowmetal}
\end{figure*}

\begin{figure*}
	\centerline{\includegraphics[width=0.92\columnwidth]{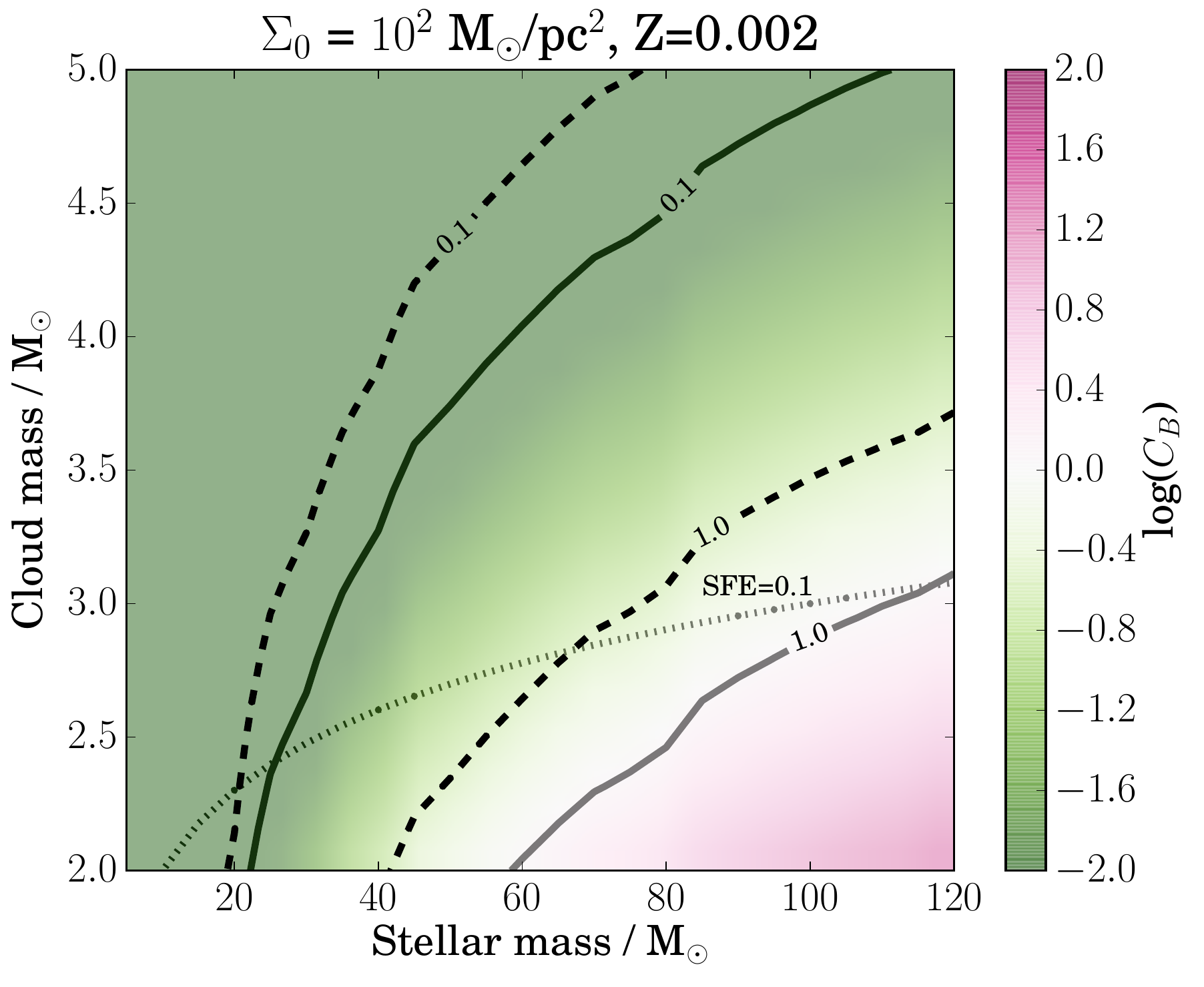}
		\includegraphics[width=0.92\columnwidth]{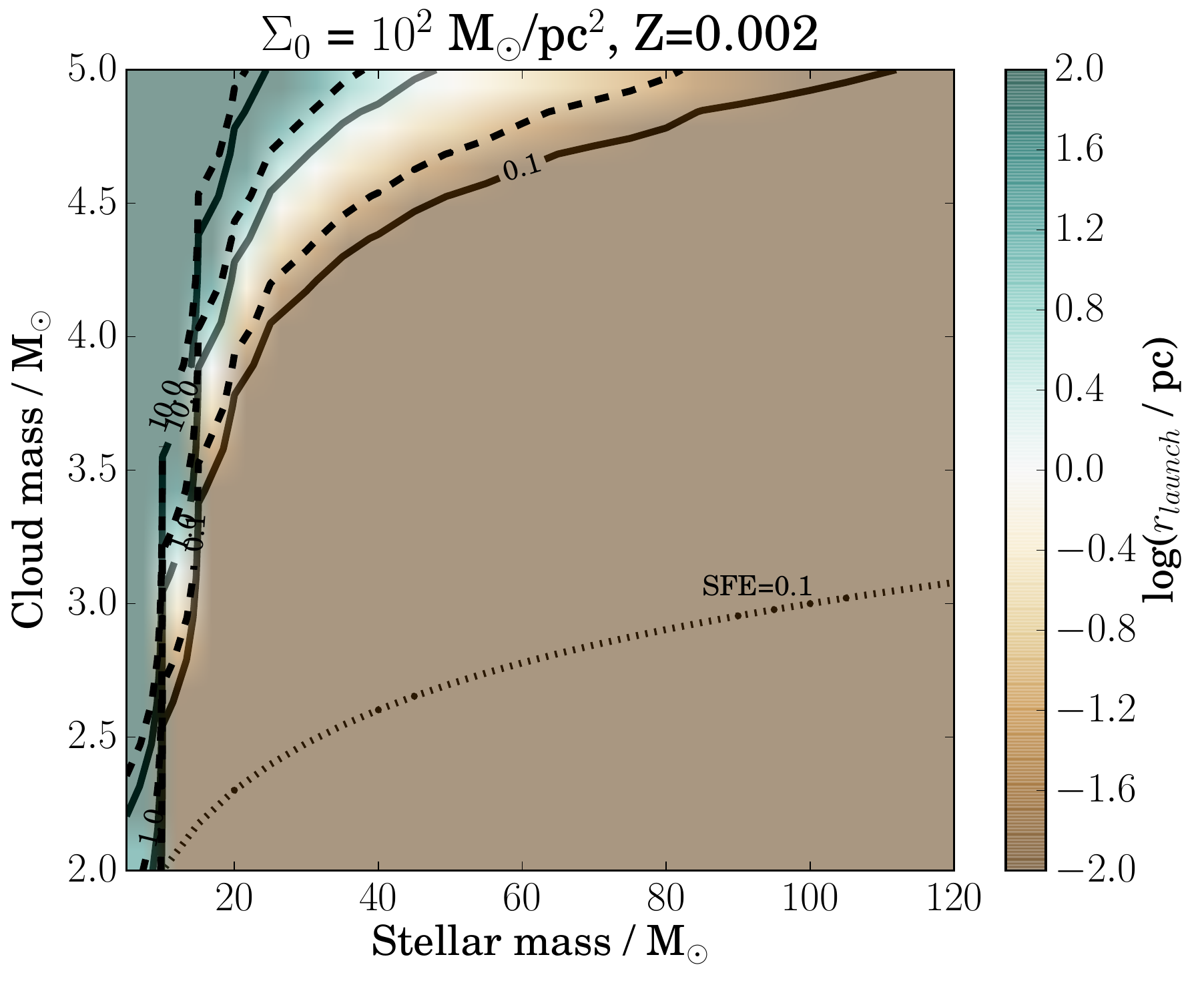}}
	\caption{Surface plots of the breakout condition $C_{B}$ (left) and the stall radius $r_{stall}$ (right) for single stars and binaries at various cloud masses at low metallicity ($Z=0.002$), enclosed by $\Sigma_0=100$ \Msolarpc. See Figures \ref{fig:rstall} and \ref{fig:breakout} for solar metallicity results.}
	\label{fig:breakoutrstall_lowmetal}
\end{figure*}

In this Section we modify the parameters used above to describe low metallicity environments. We make two changes. Firstly, we use the low metallicity stellar tracks in the Geneva tables \citep{Ekstrom2012}, where $Z=0.002$ (by comparison, solar metallicity is taken as $Z=0.014$). Secondly, the equilibrium temperature of the ionised gas is higher due to less efficient metal line cooling (see Figure \ref{fig:ionisedsoundspeed}).

In Figure \ref{fig:surface_radius_lowmetal}, we plot the criteria for winds and radiation pressure to be significant, $C_w$ and $C_{rp}$ by radius. Both parameters become smaller at low metallicity, i.e. winds and radiation pressure are less significant in these conditions. This is because firstly, winds are weaker in low metallicity stars, where the lower opacity in the stellar atmosphere reduces the ability for radiation inside the star to drive winds. Secondly, the steep dependence of the criteria on $c_i$ means that increasing $c_i$ drastically increases the efficiency of photoionisation at driving the \hii region.

In Figure \ref{fig:breakoutrstall_lowmetal}, we plot $r_{launch}$ and $C_B$ for low metallicity. Launching an accelerating \hii region is even more likely than at solar metallicity. Breakout is however less likely. This is again because the increase in $c_i$ greatly enhances the pressure inside the \hii region.

\section{Discussion}
\label{discussion}

In this Section we discuss the role of contexts outside the scope of this work on the evolution of the modelled \hii regions.

%% Already mentioned in the introduction
%\subsection{Infrared Radiation Pressure}
%
%We have neglected infrared (IR) radiation pressure in this work. \cite{Reissl2018} explored a wide range of cloud conditions more extreme than studied in this paper. The authors concluded that IR radiation pressure is likely never to be of dynamical importance due to spectral shifting of photons and higher opacity to absorption and reprocessing versus scattering. Therefore IR radiation pressure should have no impact on the results of this paper concerning the expansion of \hii regions.

\subsection{Conditions at Small Radii}

Early in the star's lifetime, the system has $r_i \rightarrow 0$. The physics of this phase is very important because, as we have established in this work, if the ionisation front does not reach the bistable point at $r_{launch}$, it cannot expand further.

The early phase of expansion of a photoionised \hii region is modelled in 1D by \cite{Keto2002}, who determine stall conditions that allow continued accretion onto the protostar. Particularly important is the fact that at small radii, the force of gravity comes mainly from the star, not the gas, as is assumed here. This means that at low radii, where the cloud mass approaches the stellar mass, our results will become less realistic and alternative formulations should be used.

Simulations by \cite{Peters2011} demostrate that disk dynamics, fragmentation and magnetohydrodynamics (MHD) play a strong role in the evolution of feedback bubbles close to the protostar, while \cite{Masson2016} and \cite{Vaytet2018} also demonstrate that non-ideal MHD is important at these scales. More recent simulation results by \cite{Kuiper2018}, which include both winds and photoionisation from an accreting protostar, argue that kinetic feedback from protostellar jets is the principal driver of the first bipolar outflows around a massive star, creating a channel for ionising radiation and stellar winds to escape.

\subsection{Conditions at Large Radii}

Once the ionisation front has left the cloud, the role of photoionisation is diminished. As \cite{Haid2018} argue, the diffuse warm \ISM is already ionised, so there is no further ionisation front. Additionally, as the cloud is dispersed, the density drops, causing the cooling time of the winds to increase. Under these conditions, the role of winds will be enhanced, and more realistic numerical solutions for the cooling of the wind bubble are required. At late times, the most massive stars die, causing the ionising photon emission rate to drop, and supernovae occur, leading to a supernova-driven structure \citep{Rahner2017} that can, under certain conditions, recollapse and lead to a second star formation event \citep{Rahner2018}. A full picture of the interaction between stars and their surroundings is important for modelling the amount of momentum and radiation injected into the \ISM of galaxies and how stars drive larger-scale flows.

The low metallicity results are important for understanding conditions at high redshift. They predict that, except in very dense conditions, photoionised \hii regions should rapidly escape star-forming regions, dispersing clouds and creating flows of a few tens of km/s. This suggests a high escape fraction from molecular clouds around the epoch of reionisation, and provides a source of low-velocity turbulence in the galactic \ISM.

\subsection{The Role of Cloud Structure}

The models in this paper consider feedback in spherical gas clumps with $1/r^2$ density profiles. Molecular clouds observed in our Galaxy are argued to have fractal structures \citep[e.g.][]{Cartwright2004}, and so our analysis only applies provided that the density field around the star can be approximated by such a density profile. The interaction between multiple \hii regions in neighbouring clumps, the ablation of clumps not containing massive stars by the \hii region, changes in the density profile and other complicating effects are omitted from this work.

At some point, \hii regions from neighbouring clumps will merge. As we discuss in Section \ref{introduction:winds}, the embedded wind bubbles are not guaranteed to merge \citep{Silich2017}, which will reduce their efficiency, while leakage from porous shells further reduces their efficiency \citep{Harper-Clark2009}. This will counter the effect in Equation \ref{wind:coefficient}, where winds become more effective the more stars are emitting inside a single wind bubble and \hii region.

\subsection{Observational Significance}
\label{discussion:obs}

\rev{The structure of \hii regions affects their observable properties, since the density, temperature and radiation field determine the emission of radiation from the gas at each position inside the region. It is thus very important to understand the combined behaviour of photoionisation, winds and radiation pressure if we are to successfully interpret observed regions and understand the underlying physics of real systems. For example, models of photoionisation-driven regions have recently been used to interpret the expansion of nearby \hii regions in \cite{Tremblin2014a} and \cite{Didelon2015}. In particular, the latter paper makes it clear that the underlying assumptions about the region and its behaviour strongly shape our interpretation of its current evolutionary state. If the expansion rate is faster than we predict, then we will overpredict the age of the source.}

\rev{In Sections \ref{wind:ratioofbubbles} and \ref{wind:density-of-photoionised-gas-and-observational-significance}, we argue that even in cases where winds do not have a strong influence on the expansion of the \hii region, a large wind bubble can exist. Such wind bubbles are observed in X-ray emission, since high gas temperatures are required to produce such photons, but they also affect the shape and structure of the \hii region, which forms a shell between $r_w$ and $r_i$. We further note in Section \ref{radiation_pressure:strong} that if radiation pressure's influence is large enough, the combined effect of radiation pressure and winds on the photoionisation solution becomes complex and can no longer be treated as a perturbation. Having identified likely regions of parameter space where this occurs, we must turn to more complete calculations to provide these solutions.}

\rev{More complete analytic models of this nature have been used to target specific regions in \cite{Pellegrini2007} and \cite{Pellegrini2009}. \cite{Draine2011} focuses on parts of the parameter space where radiation pressure has a significant effect on the density structure of the photoionised gas, which in turn shapes the emission from the \hii region. One quantity used to interpret such emission from \hii regions is the ionisation parameter, i.e., the ratio of ionising photon to hydrogen density. This quantity has been calculated by \cite{Yeh2012} and \cite{Pellegrini2012} using full 1D hydrostatic analytic models including winds and radiation pressure. Emission from individual lines is also a useful tracer, and has been computed similarly in \cite{Yeh2013} and \cite{Verdolini2013}. Such approaches are highly successful in reproducing emission line ratios from observed regions in nearby galaxies \citep{Pellegrini2019}. Complete radiative transfer models are also essential to obtain an accurate picture of the interaction between radiation at various energies and the gas inside \hii regions.}

%\rev{1D models are a useful tool for computing large parameter spaces due to their low computational cost compared to full 3D hydrodynamic simulations. However, certain features of observed \hii regions depart from this 1D structure, such as clump fragmentation and evaporation, merging of regions inside a single giant molecular clouds, escape of pressurised gas through low-density channels in the cloud, full hydrodynamic turbulence and other phenomena. While models exist to approximate these various factors in the works cited in this paper, there are certain features that require full 3D simulations. Nonetheless, 1D models remain important for their simplicity and predictive power.}

\section{Conclusions}

In this paper, we a) develop algebraic models to describe the expansion of photoionised \hii regions under the influence of gravity and accretion in power-law density fields, b) determine when terms describing winds, radiation pressure, gravity and photon breakout become significant enough to affect the dynamics of the \hii region, and c) solve these expressions for a set of physically-motivated conditions. We provide a set of expressions that can be used to calculate the dynamics and structure of observed \hii regions, particularly the transition from thin shell \hii regions to volume-filling \hii regions and the role of radiation pressure in setting internal density gradients.

Photoionisation heating is very efficient at driving an ionisation front into cloud with steep power-law density fields (index $w > 3/2$). The timescale for this expansion is set by the free-fall time in the cloud. In uniform density fields, ionisation fronts tend towards a ``stall'' radius where the expansion rate tends to zero. If $w > 3/2$, this is replaced with a ``launch'' radius where the ionisation front accelerates to infinity if the initial radius is larger than the launch radius.

We then focus on a \rev{singular} isothermal sphere ($w=2$) and introduce winds, radiation pressure and photon breakout, as well as calculating the launch radius. As wind bubbles cool efficiently in dense cloud environments, both winds and radiation pressure are most important when the \hii region radius is small ($\sim$ 0.1 pc radius). Their importance is enhanced at high surface densities. \rev{Under a larger range of conditions, their effect on the dynamics of the \hii region is around 10\% of the contribution from photoionisation, although radiation pressure can have a larger effect over a greater range. We caution that radiation pressure is a complex phenomenon. The radiation pressure parameters given in this paper are fiducial, but microphysics could decrease or even increase their values. The goal of this work is to provide estimates of which conditions it has a significant influence in, so that more detailed work can be carried out.}

Both winds and radiation pressure shape the internal structure of the \hii region even when they are not dynamically important, which has observational consequences. Conversely, a large wind bubble does not automatically mean that winds are dynamically important. \rev{However, winds, radiation pressure and photoionisation interact in non-linear ways, so even when winds or radiation pressure are not the most important process, care should be taken to consider their influence on the \hii region once they approach non-negligible levels.}

At solar metallicity, some photon breakout is possible for high mass stars in low density environments, while gravity prevents the expansion of \hii regions only at very high densities. The calculations for photon breakout are lower limits, since this work omits shell fragmentation and non-spherical cloud geometries.

At low metallicity, photoionisation becomes more efficient due to the higher equilibrium temperature of the photoionised gas. The role of winds, radiation pressure, photon breakout and gravity are all diminished. This has implications for galaxy formation simulations and theory, including the unresolved influence of photoionisation-driven feedback and higher escape fractions than might be assumed when the structure of molecular clouds around young massive stars is unresolved.

The broad picture given by this paper is that \hii regions around young massive stars are first driven by winds, radiation pressure or some other dynamical effect up 0.01 to 0.1 pc. Above this radius, photoionisation accelerates the ionisation front out of the cloud, where it enters the warm interstellar medium or interacts with other dense clumps inside the star-forming region. This picture is complicated by the precise cooling rate of stellar winds, \rev{the detailed microphysics affecting radiation pressure inside \hii regions, and conditions} at high densities such as the Central Molecular Zone where gravity, winds, and radiation pressure compete with photoionisation.

\section*{Acknowledgements}
The authors would like to thank the referee for their careful and thoughtful comments in improving the paper. The authors would also like to thank Ashley Barnes, Simon Glover, Jo Puls and Daniel Rahner for their useful comments and discussions during the preparation of this work. We acknowledge funding from the European Research Council under the European Community's Seventh Framework Programme (FP7/2007-2013). SG and RSK have received funding from Grant Agreement no. 339177 (STARLIGHT) of this programme. EWP and RSK further acknowledge support from the Deutsche Forschungsgemeinschaft in the Collaborative Research Centre SFB 881 ``The Milky Way System'' (subprojects B1, B2, and B8) and in the Priority Program SPP 1573 ``Physics of the Interstellar Medium'' (grant numbers KL 1358/18.1, KL 1358/19.2). SG acknowledges support from a NOVA grant for the theory of massive star formation.

%%%%%%%%%%%%%%%%%%%%%%%%%%%%%%%%%%%%%%%%%%%%%%%%%%



\bibliographystyle{mnras}
\bibliography{samgeen} % if your bibtex file is called example.bib

%%%%%%%%%%%%%%%%%%%%%%%%%%%%%%%%%%%%%%%%%%%%%%%%%%

%%%%%%%%%%%%%%%%% APPENDICES %%%%%%%%%%%%%%%%%%%%%

\appendix

\section{Pressure Velocities in and Isothermal Cloud}
\label{appendix:pressure_velocity}

\rev{In this paper we define external forces on the shell in terms of velocity $v_0$, in order to provide a direct correction to the expansion velocity $r_i$. }

\rev{We define $v_0$ as}
\begin{equation}
v_0^2 = 2 f G M(<r) r^{-1} = f v_{esc}^2
\label{photo:v0_generic}
\end{equation}
\rev{where $v_{esc}$ is the escape velocity at $r$. $f$ is a scaling factor that we determine below. For a cloud where $w=2$, $f$ is constant with respect to radius and on the order of unity except in the case where a central mass (i.e. the stellar source) exerts more gravitational force than the gas density $M(<r)$.}

\rev{We assume in this paper that accretion onto the shell is modelled by a ram pressure, where the gas is at density $\rho_0$ and velocity $v_{esc}$. This gives $f=1$. }

\rev{For gravitational acceleration, we take $v_0$ to be represented by the orbital velocity $\sqrt{G M(<r) r^{-1}}$. This gives $f=1/2$. The combined $f$ including both terms is thus $3/2$.}

\rev{Note that, as discussed in Section \ref{cloud}, we ignore the time evolution of $\rho_0$ due to accretion and the effect of the central mass of the star or cluster. This changes the form of Equation \ref{photo:v0_generic} and introduces radial dependencies to $v_0$. For examples of these calculations, see, e.g. \cite{Keto2002} and \cite{Didelon2015}.}

\section{Expansion of a Photoionisation Region with External Pressure}
\label{appendix:stalled_photoionisation}

In Section \ref{photoionisation_only} we give Equation \ref{photo:dynamical_equation} to describe the expansion of an ionisation front with an external pressure term, described as a velocity $v_0$ (see Appendix \ref{appendix:pressure_velocity}).

We begin with Equation \ref{cloud:profile} that describes the density profile in the neutral gas at $r$:
\begin{equation}
n(r) = n_0 (r / r_0)^{-w},
\end{equation}
Equation \ref{photo:equilibrium} that describes photoionisation equilibrium inside the ionisation front $r_i$:
\begin{equation}
\frac{4 \pi}{3} r_i^3 n_i^2 \alpha_B = Q_H,
\end{equation}
and Equation \ref{photo:externalpressure} that describes pressure equilibrium at $r_i$:
\begin{equation}
n_i c_i^2 = n(r_i) (\dot{r}_i + v_0)^2.
\end{equation}

Equation \ref{photo:externalpressure} can be written in terms of density squared as
\begin{equation}
n_i^2 c_i^4 = n(r_i)^2 (\dot{r_i} + v_0)^4
\end{equation}
Substituting Equation \ref{photo:equilibrium} for $n_i$ and Equation \ref{cloud:profile} for $n(r_i)$, this gives
\begin{equation}
c_i^4 \frac{Q_H}{\alpha_B}\frac{3}{4 \pi} = n_0^2 r_0^{2 w} r_i^{3-2 w} (\dot{r}_i + v_0)^4
\end{equation}
When $r_i = r_{stall}$, $\dot{r}_i = 0$. We can thus divide the above equation by itself when these two terms are substituted:
\begin{equation}
1 = \left( \frac{r_i}{r_{stall}} \right)^{3-2w} \left( \frac{\dot{r}_i + 1}{v_0} \right)^{4}
\end{equation}

Hence if $R \equiv r_i / r_{stall}$, we recover Equation \ref{photo:dynamical_equation}. This gives us:
\begin{equation}
r_{stall}^{4 w - 7} = \left ( \frac{3}{4 \pi} \frac{Q_H}{\alpha_B} \right )^{-1} \left ( 8 \pi f G \frac{m_H}{X} \right )^2 \left ( n_0 r_0^2 \frac{1}{c_i} \right )^4 
\label{appendix:photo:rstall}
\end{equation}
as in \cite{Geen2015b}.

We can define a typical freefall time for the cloud
\begin{equation}
t_{ff,0}^2 = \frac{3 \pi}{32 G \rho_0}
\end{equation}
which using Equations \ref{cloud:mass} and \ref{photo:v0_generic} \rev{where $w=2$} gives
\begin{equation}
v_0 = - \frac{r_0}{t_{ff,0}} \frac{\pi}{2} \sqrt{3f}
\label{appendix:pressure_velocity_v0tff}
\end{equation}

Substituting for $v_0$ using Equation \ref{appendix:pressure_velocity_v0tff} and taking $w=2$ gives
\begin{equation}
\dot{R} = R_0 \frac{\pi}{2} \frac{\sqrt{3f}}{t_{ff,0}}(R^{1/4} - 1)
\end{equation}
where $R_0 \equiv r_0 / r_{stall}$. As stated previously in the paper, for $w>3/2$, $r_{stall}$ becomes a ``launching'' radius $r_{launch}$ that the solutions accelerate away from rather than converging to.

% REMOVED BECAUSE RSTALL IS MORE ACCURATE
%\section{Stall Radius and Characterisation}
%\label{appendix:rstall}
%
%If the initial state of the \hii region is such that $r < r_{stall}$, the ionisation front is crushed. Traditional methods of integrating out to the Str\"omgren radius in a uniform medium (i.e. the radius at which the initial density field reaches ionisation equilibrium) do not work, since the calculation in an isothermal cloud contains an asymptote. Instead, we note the pressure balance
%\begin{equation}
%\frac{2}{\gamma} n_i c_i^2 = n(r_i,t) (\dot{r}_i + v_0)^2
%\end{equation}
%if $n_i = n(r_i,t)$ before the density profile in the \hii region has begun to evolve, we have the condition for stalling 
%\begin{equation}
%\frac{2}{\gamma} c_i^2 = v_0^2
%\end{equation}
%
%In other words, whether or not the ionisation front enters the crushed state depends solely on this balance between the sound speed in the ionised gas and the pressure terms acting on the ionisation front \citep[an argument also made by][]{Dale2012}.

\section{Wind and Photoionisation Dynamical Equations}
\label{appendix:wind}

In this Section we derive Equation \ref{wind:dynamics} that describes the dynamical evolution of an ionisation front with an embedded wind source. 

The solution is similar to that in Equation \ref{appendix:stalled_photoionisation}. We amend Equation \ref{photo:equilibrium} to include a collisionally-ionised sphere up to $r_w$:
\begin{equation}
\frac{4 \pi}{3} (r_i - r_w)^3 n_i^2 \alpha_B = Q_H.
\label{appendix:wind:photoequilibrium}
\end{equation}

In addition, we have Equation \ref{wind:windpressurebalance_momcons} that describes the pressure balance at $r_w$ from a momentum-conserving wind:
\begin{equation}
\frac{\dot{p}_w}{4 \pi r_w^2} = n_i c_i^2 \frac{m_H}{X}.
\label{appendix:wind:balance_momcons}
\end{equation}

We begin by substituting $r_w$ in Equation \ref{appendix:wind:photoequilibrium} with Equation \ref{appendix:wind:balance_momcons}:
\begin{equation}
r_i^3 - \left ( \frac{\dot{p}_w}{4 \pi} {c_i^2}\frac{X}{m_H}\frac{1}{n_i} \right )^{3/2} = \frac{Q_H}{\alpha_B} \frac{3}{4 \pi} \frac{1}{n_i^2}
\end{equation}

We then substitute $n_i$ in this equation for Equation \ref{photo:externalpressure}
\begin{equation}
\begin{split}
r_i^3 - \left ( \frac{\dot{p}_w}{4 \pi}\frac{X}{m_H} \right)^{3/2} \left(\frac{n_0 r_0^2(\dot{r_i} + v_0)^2}{r_i^2} \right )^{-3/2} = \\ \frac{Q_H}{\alpha_B} \frac{3}{4 \pi} \left ( \frac{1}{c_i^2}\frac{n_0 r_0^2 (\dot{r_i} + v_0)^2}{r_i^2}  \right )^{-2}
\end{split}
\end{equation}
Multiplying and dividing by terms in $r_i$ and $\dot{r}_i + v_0)$ we obtain
\begin{equation}
(\dot{r_i} + v_0)^4 = \left ( \frac{\dot{p}_w}{4 \pi}\frac{X}{m_H} \frac{1}{n_0 r_0^2}\right )^{3/2} (\dot{r_i} + v_0) + \frac{Q_H}{\alpha_B}\frac{3}{4 \pi}  \left ( \frac{c_i^2}{n_0 r_0^2} \right )^{2} r_i
\end{equation}
which gives us Equation \ref{wind:dynamics} and terms for $A_w$ and $A_i$ as defined there.

\section{The Role of Gravity in the \hii Region}
\label{appendix:gravityinthehiiregion}

In addition to radiation pressure, self-gravity can also set up a gradient inside the \hii region. We assume here that the \hii region has a uniform density $n_i$ and calculate the conditions where the pressure from gravity should set up a gradient. 

The pressure from self-gravity acting on a spherical shell of thickness $dr$ at radius $r$, where $r < r_i$ is
\begin{equation}
dP_g = -\frac{G M_i(<r)dm}{r^2} \frac{1}{4 \pi r^2},
\end{equation}
where $M_i(<r)$ is the mass in ionised gas contained within $r$, or $\frac{4}{3} \pi r^3 n_i \frac{m_H}{X}$, the mass of the spherical shell $dm=\frac{m_H}{X}n_i 4 \pi r^2 dr$. This gives
\begin{equation}
dP_g = - 4 \pi G n_i^2 \left( \frac{m_H}{X} \right)^2 \frac{r}{3} dr.
\end{equation}

Equating this to the thermal pressure of the photoionised gas at $r$ and integrating the result, we obtain
\begin{equation}
\frac{n_{inner}}{n_{outer}} = C_{g} + 1
\end{equation}
where
\begin{equation}
C_g \equiv \frac{\pi G r^2 m_H}{6 c_i^2 X} n_{inner}.
\end{equation}
Note that the condition $C_g \ll 1$ defines whether or not the gradient due to gravity is important, since these solutions assume that $n_{inner} \simeq n_{outer}$. In Equation \ref{wind:density_contrast} we argue that $n_i / n(r) \simeq 0.1$ to 1. We can thus substitute for $n_i r^2 \simeq n(r) r^2 = n_0 r_0^2$. Substituting this with Equations \ref{cloud:r0} and \ref{cloud:n0}, we can write
\begin{equation}
C_g = 6.37\times10^{-4}  \left(\frac{c_i}{10 \mathrm{km/s}}\right)^{-2}
\left(\frac{M_0}{100 \mathrm{M}_{\odot}}\right)^{1/2}
\left(\frac{\Sigma_0}{100 \mathrm{M}_{\odot}\mathrm{pc}^{-2}}\right)^{1/2}
\end{equation}
Since this is an upper bound, and requires very dense clouds with low temperature photoionisation regions, we neglect the effect of self-gravity inside the \hii region in this work.

\bsp	% typesetting comment
\label{lastpage}
\end{document}